\documentclass[english]{article}
\usepackage[T1]{fontenc}
\usepackage[latin9]{inputenc}
\usepackage{amsmath}
\usepackage{graphicx}
\usepackage{longtable,booktabs}
\usepackage{graphicx,grffile}

\def\re#1{(\ref{#1})}   
\def\dnum{597} 
\providecommand{\tabularnewline}{\\}
\newcommand{\lyxdot}{.}

\newcommand{\lyxaddress}[1]{
\par {\raggedright #1
\vspace{1.4em}
\noindent\par}
}

\makeatother

\usepackage{babel}
\begin{document}

\title{Silence measurements and measures for ET: characterisation of long term seismic noise  in the Mátra Mountains}

\author{Somlai L.$^{1,2}$, Gráczer Z.$^{3}$, Vasúth M.$^{1}$, Wéber Z.$^{3}$,
Ván P.$^{1,4,5}$}
\maketitle

\lyxaddress{$^{1}$MTA Wigner Research Centre for Physics, Institute of Particle and Nuclear Physics, 1121 Budapest, Konkoly Thege Miklós út 29-33, Hungary;\\
$^{2}$University of Pécs, Faculty of Sciences , H-7624 Pécs, Ifjúság út 6, Hungary \\
$^{3}$MTA Research Centre for Astronomy and Earth Sciences, Geodetic and Geophysical Institute, H-9400, Sopron, Csatkai E. u. 6-8, Hungary; \\
$^{4}$Department of Energy Engineering, Budapest University of Technology and Economics, 1111 Budapest, Bertalan Lajos u. 4-6, Hungary; \\
$^{5}$Montavid Thermodynamic Research Group, Budapest, Hungary.}

\begin{abstract}
The analysis of long term seismological data collected underground in the Mátra Mountains, Hungary, using the facilities of the Mátra Gravitational and Geophysical Laboratory (MGGL) is reported. The laboratory is situated inside the Gyöngyösoroszi mine, Hungary, 88m below the surface. This study focuses on the requirements of the Einstein Telescope (ET), one of the planned third generation gravitational wave observatories, which is designed for underground operation. After a short introduction of the geophysical environment the evaluation of the collected long term data follows including the comparison of a two-week measurement campaign deeper in the mine. Based on our analysis and considering the specialities of long term data collection, refinements of the performance and evaluation criteria are suggested as well as performance estimation of a possible Mátra site.
\end{abstract}

\section{Introduction}

The improved sensitivity of the future third generation gravitational wave (GW) detectors requires various technological developments. One of the plans is to optimize the facility for underground operation, in order to reduce the noise between the frequencies from 1~Hz to 10~Hz, because according to the related sensitivity calculations the seismic noise and the Newtonian-noise result the most important contribution in this frequency range \cite{BraEta10p,Har15a}. During the preparatory studies of the so-called Einstein Telescope (ET), the European initiative, several short term seismic measurements were performed in various locations \cite{ETdes11r,BekEta15a,Bek13t}. Based on these studies two seismic noise related criteria were established \cite{BekEta15a}: a spectral and a  cumulative. According to the spectral criteria the average horizontal acceleration Amplitude Spectral Density (AASD) must be smaller than 
\begin{equation}
A_{BF}=A^{(a)}(\,f\,)=2\cdot10^{-8}\frac{m/s^{2}}{\sqrt{Hz}},\,\,\:\text{if}\,\, 1Hz\leq f\leq 10Hz.\label{eq:ETReqBF}
\end{equation}

This is the so called Black Forest line, named after one of the investigated sites. This spectral requirement corresponds to a cumulative value, the square root of the displacement Power Spectral Density (PSD) integrated from the Nyquist frequency down to 2~Hz. This is the 2~Hz root mean square ($rms_{2Hz}$) and the criteria is its compatible value with he Black Forest line, $0.1\,nm$. 
In the ET survey the three best sites that fulfilled these requirements are the LSC Canfranc laboratory in Spain ($rms_{2Hz}=0.070\,nm$), the Sos Enattos mine in Sardinia, Italy ($0.077\,nm$) and the Gyöngyösoroszi mine in Hungary ($0.082\,nm$ and  $0.12\,nm$ in depths 400m and 70m respectively). The data collection was performed for maximum one week periods in the various sites. It was established also that the depth of the sites is not a primary factor (Fig. 3.11 in \cite{Bek13t}), the local population density matters, and the type of the rock mass is important: hard rocks are less noisy. 

The Mátra Gravitational and Geophysical Laboratory (MGGL) is operating since March 2016 with the purpose to evaluate and survey the Mátra mountain range as a possible ET candidate site. The primary goal of the laboratory is to collect seismic noise data for a longer period and evaluate them according to any requirements of ET. Beyond seismometers several other instruments are registering infrasound and electromagnetic noise and also detecting muons here \cite{BarEta17a,BarEta16am}. The laboratory is located in the coordinates (399 MAMSL, 47$^{\circ}$52' 42.10178\textquotedbl{}, 19$^{\circ}$51' 57.77392\textquotedbl{} OGPSH 2007 (ETRS89)), along a horizontal tunnel of the mine, 1280~m from the entrance, 88~m depth from the surface. It is situated near to the less deep location of the above mentioned former short term measurements and it is prepared for long term automatic online data collection. In the laboratory a Guralp CMG-3T seismometer (hereafter referred as ET1H) is installed and was operating continuously except shut downs and interfering mine activities (e.g. explosions). We have also performed a measurement campaign for two weeks, when one more identical seismometer was installed in a deeper location, 3764~m from the entrance of the  horizontal tunnel and 404 m below the surface, very close to the second location of the previous short measurement of Beker, too. 

There is an ongoing re-cultivation in the mine and therefore the human activity is not negligible. The regular operation of the mine railway, the continuously working large water pumps in the vicinity of the laboratory and the related technical service and construction activities are producing human noise that would not be present during gravitational wave detection. A part of the surface originated civilization noise and the general seismological activity is unavoidable as well as the seasonal variation of noise due to weather and vegetation conditions. In principle our collected data makes possible estimate, compare and analyse these various noise sources. However, their effect were surprisingly unpredictable in the investigated frequency range and the role of not identified noises was far more important in the long term balance, therefore their specific direct filtering  proved to be meaningless. Moreover, in case of long term measurements a filtering of the data could bias the performance measures and the different sites could not be compared. 

Therefore this paper has two aims. We report the long term ET related noise performance of the Mátra site and also investigate various aspects for possible  refinement of performance characteristics and methodology in order to better understand the overall, long term detection possibilities of gravitational waves in an underground location. The particular aspect of the required analysis is the presence of various short term seismological and seismic disturbances with large amplitudes. These are unpredictable, unavoidable and must be filtered to obtain reliable estimation of the average low noise level. However, any particular truncation or cutting biases and distorts the spectral and also the cumulative noise measures. {\em Therefore  we suggest to use the percentiles of the unfiltered data for this purpose.} 

The evaluation of spectra and $rms$ requires an intermediate averaging over the short term -- 128s or 50s long -- averages. We suggest to use the daily percentiles of the unfiltered data. The percentiles filter the highest and lowest values and this filtering is relative, intrinsic to the data set. The daily averages consider the natural periodicity of the data instead of some arbitrary values, like the 14x128s averaging of Beker {\it et al} \cite{BekEta15a}. In particular we use the averaged daily median values for the performance levels. Also the averaged daily 10th and 90th percentiles of the unfiltered data can be used to estimate the spectral and cumulative variation of the data set. 

The paper is organized as follows. First we shortly introduce the geological and seismological environment. Then we give a detailed description of the evaluation procedure including some sensitive aspects, like the usage of modus for $rms$ calculation. In the fourth section we analyse the effect of averaging on spectral and cumulative measures established by Beker {\it et al}. Then we investigate daily and seasonal changes and also the effect of depth with the data collected from a deeper location during a two-week measurement campaign. Finally we summarize performance characteristics of the Mátra site and emphasize the most important methodological aspects for similar future studies.

\section{The Mátra mountain range, the geophysical environment of MGGL}

\begin{figure}
\centering
\includegraphics[width=0.8\textwidth]{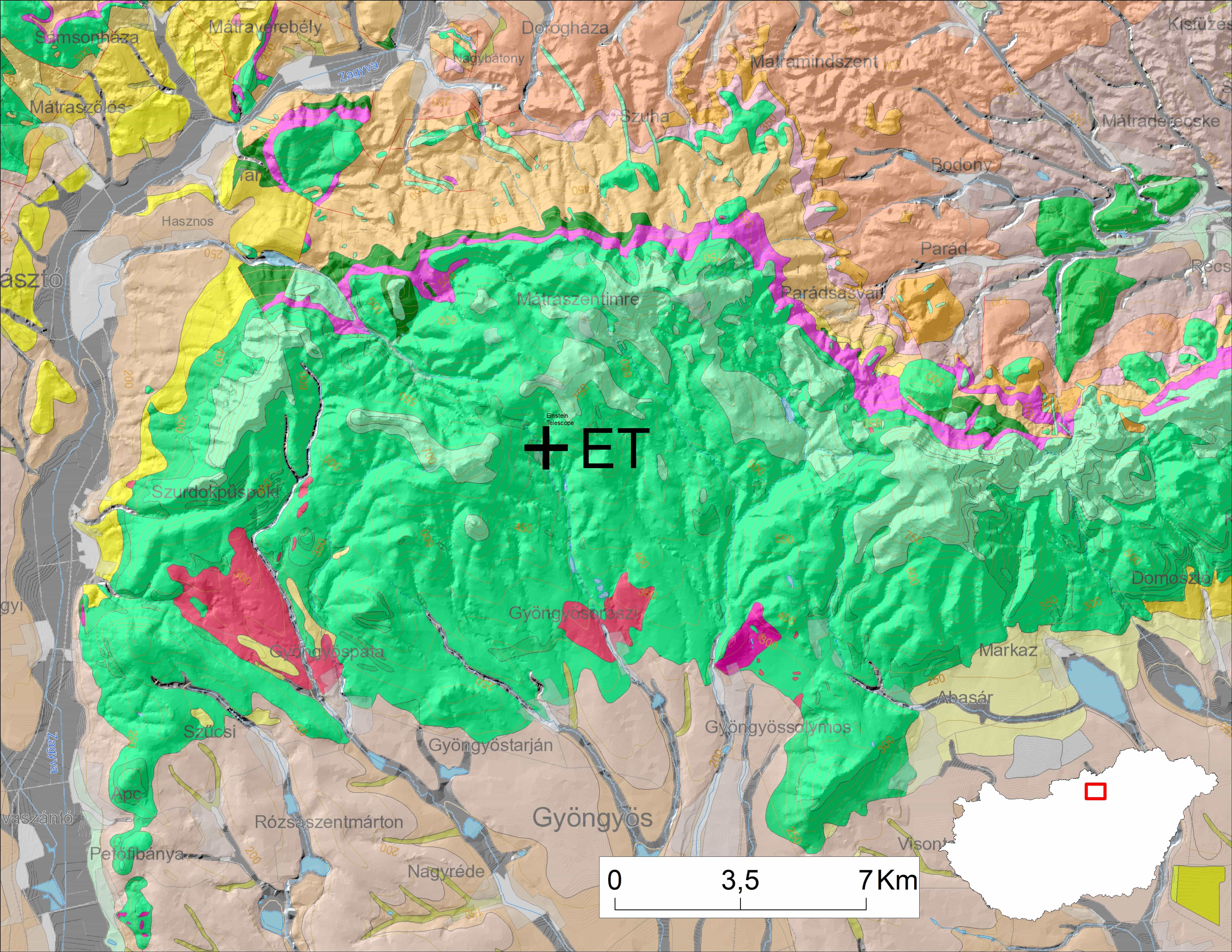}
\caption{The surface lithology of the Mátra area. The green part denotes various andesitic rocks. {\small\it(The map is composed by Tamás Biró using https://map.mbfsz.gov.hu/fdt100/)}}\label{Matrand}
\end{figure}

The lithological composition of Mátra mountain range is moderately blocky andesite, as it is shown in Figure \ref{Matrand}, where the green areas denote various andesite types born at the same geological era \cite{KovSza08a}. The most important physical properties of a typical hard rock from Gyöngyösoroszi mine are given in Ref. \cite{BarEta17a}. In the following we characterise the seismicity and the noise from nearby explosions in the surrounding area. In order to enhance comparison with other sites we also show quantitative measures by calculating the related ground displacement nd seismic hazard of the Mátra Mountains.

\subsection{Seismicity of the Mátra Mountains and the surrounding areas}\label{seismicity-of-the-matra-mountains-and-the-surrounding-areas}

\begin{figure}
\centering
\includegraphics{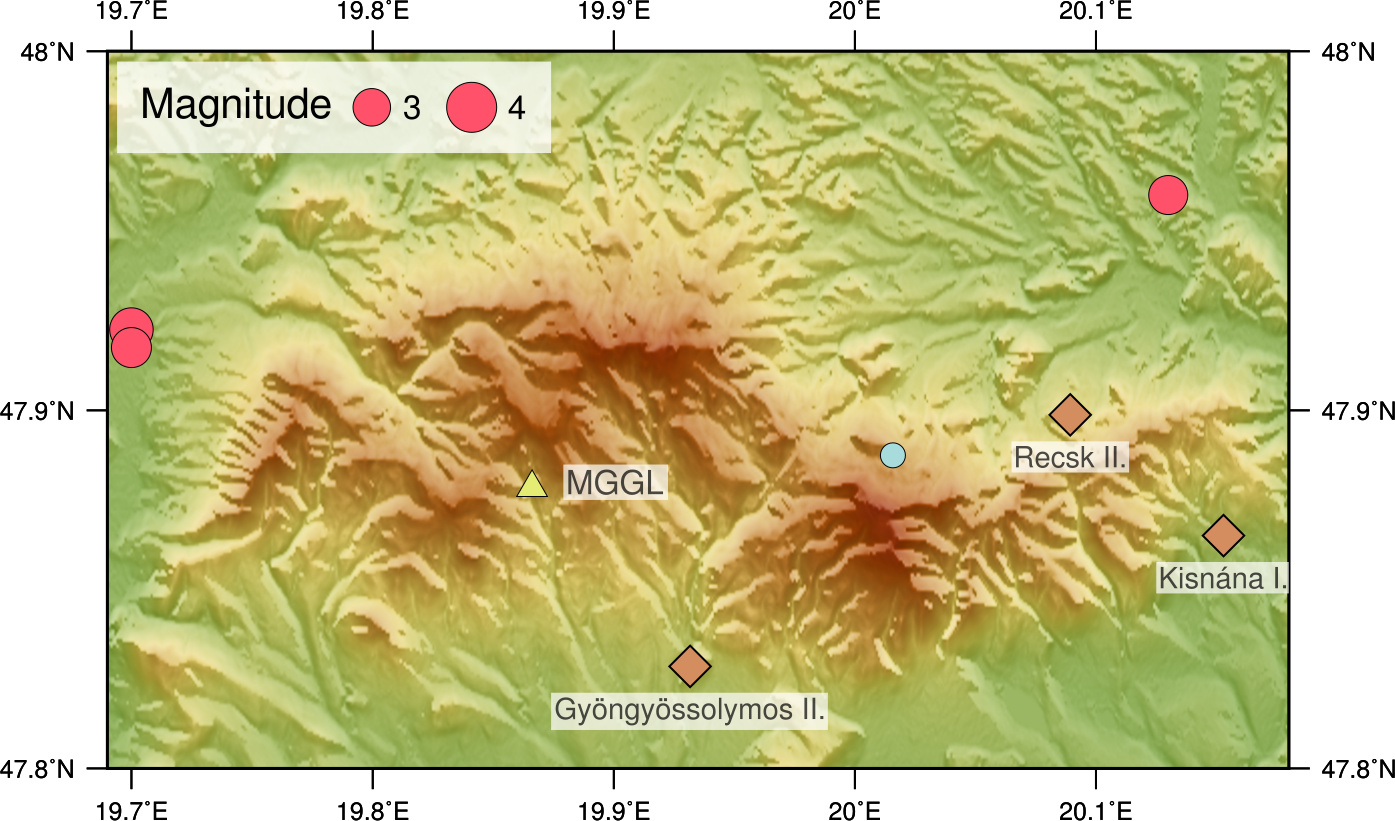}
\caption{The Mátra area. The triangle shows the location of the MGGL, while the diamonds mark the position of the most active quarries. Red circles: known earthquakes in the area, blue circle: explosion of unknown magnitude.}
\label{fig_earthquake}\end{figure}

The level of seismic activity in the Mátra Mountains is low. Figure \ref{fig_earthquake} shows the epicentres of the known earthquakes of the Mátra area (19.69-20.18E, 47.8-48.0N) based on the data of the Hungarian National Earthquake Catalogue and Hungarian National Seismological Bulletins. Only earthquakes with magnitude greater than 3.0 are shown as lower magnitude events can be misclassified quarry explosions. It can be seen that only three small earthquakes (M\textless{}=3.5) were ever observed in the area in 1879, 1895 and 1980.

The seismicity of the Carpathian basin as a whole can be considered moderate. The known earthquakes of the area with magnitude larger than 3.0 are shown in Fig. 3. According to the collected data on average one earthquake with magnitude \(M\ge5.0\) can be observed in the region annualy.

\subsection{Quarry explosions in the Mátra Mountains}\label{quarry-explosions-in-the-matra-mountains}

Official information on the open pit mine explosions in the Mátra Mountains is available for three quarries : Gyöngyössolymos II., Kisnána I. and Recsk II. According to the reports 91 explosions have been detonated in these mines between March 2016 and December 2017. Most of them - 52 - were performed in the quarry of Kisnána. The number was somewhat smaller - 34 - in Recsk and only 5 blasts were carried out in the Gyöngyössolymos mine.

\begin{figure}
\centering
\includegraphics[width=\textwidth]{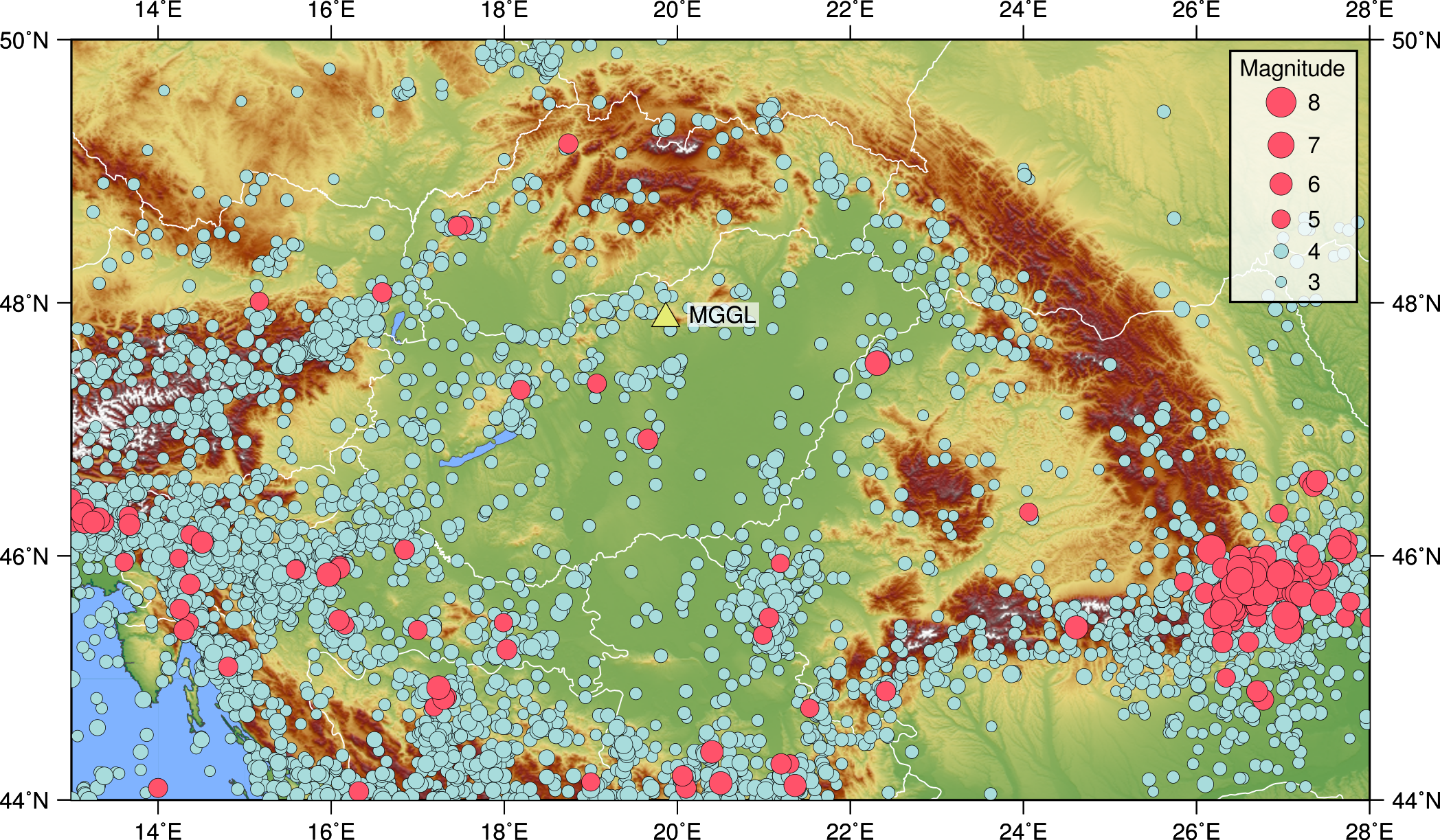}
\caption{Seismic activity in the Carpatho-Pannonian area. The triangle shows the location of the MGGL. The circles mark the epicentres of the known earthquakes occurred after the year 1800. Blue circles: earthquakes with magnitudes 3.0\textless{}=M\textless{}5.0, red circles: earthquakes with magnitudes M\textgreater{}=5.0.}
\end{figure}

We have tried to identify these events on the data registered by the ET1H seismological station deployed in the MGGL. It must be noted that for technical reasons the station was did not operate in the periods 2016-09-08 - 2016-09-19, 2016-10-18 - 2016-11-17, 2017-11-26 - 2017-11-30, 2017-12-15 - 2017-12-31 (altogether 65 days). If an event occurred in one of these periods, we tried to find its signal on the data of PSZ station, which is located on the surface, 4.88km from MGGL. The process of identification was hampered by the fact that in many cases the mines reported only the date of the blasts and not the time and even when the time was indicated it often deviated from the actual time by several minutes. Eventually however we have successfully identified 85 events on the seismograms.

The MTA CSFK GGI Kövesligethy Radó Seismological Observatory is continuously monitoring the seismicity of Hungary and the surrounding territories. Mine explosions are very often detected, however for the reliable location of a seismic event data from multiple stations are needed, therefore the focal parameters of small energy blasts can not be determined. During the processing of the data of the Hungarian National Seismological Network \cite{GraEta18a} and the stations of the surrounding countries for the years 2016-2017 we have localised 20 events in the Mátra area. The magnitude of the largest one was M=2.1. These events were identified as quarry blasts with only one exception. This event, based on its characteristics, was also a mine explosion, however its origin remained unknown.

\subsection{Ground displacements caused by seismic
events}\label{ground-displacements-caused-by-seismic-events}

We have selected four events to characterise the ground displacement in the MGGL caused by different seismic events. For the computations of displacements 100 sps streams were used. The data before the instrument correction were detrended and filtered by a second order high pass Butterworth filter with a corner frequency of 1 Hz.

As the seismicity in the Mátra area is mainly determined by the mining activities in the quarries, we computed the ground displacement for characteristic explosions carried out in the three most active mines. Table 1 shows the maximum displacement amplitudes for the three components in the case of the selected events. The magnitude of the displacement is similar for the mines Gyöngyössolymos and Kisnána, while the maximum displacement for the explosion belonging to Recsk mine is an order of magnitude smaller.

As the recent Tenk earthquake (2013-04-22, M=4.8, epicentral distance=42 km) occurred relatively close to the MGGL it may be of interest. Unfortunately, at the time of the earthquake the ET1H station was not yet installed. However we can use the data of the PSZ station which was operating during that period. The fourth row of the Table 1. shows the displacements observed at the PSZ station. It can be seen that the Tenk earthquake produced around 20-30 times larger displacements than the strongest explosion. It must be noted that in the MGGL - due to the large subsurface depth - this value probably would have been smaller.

\begin{longtable}[]{@{}llrrr@{}}
\caption{Observed maximum ground displacement values for frequencies
larger then 1 Hz}\tabularnewline
\toprule
Station & Event & Z {[}\(\mu m\){]} & N {[}\(\mu m\){]} & E
{[}\(\mu m\){]}\tabularnewline
\midrule
\endfirsthead
\toprule
Station & Event & Z {[}\(\mu m\){]} & N {[}\(\mu m\){]} & E
{[}\(\mu m\){]}\tabularnewline
\midrule
\endhead
ET1H & 2016-05-10 10:47 (explosion; Gyöngyössolymos; dist= 7.4 km) &
0.88 & 0.66 & 0.53\tabularnewline
ET1H & 2017-05-30 10:43 (explosion; Kisnána; dist=21.5 km) & 0.10 & 0.07
& 0.08\tabularnewline
ET1H & 2017-06-22 11:21 (explosion; Recsk; dist=16.8 km) & 0.34 & 0.78 &
0.43\tabularnewline
PSZ & 2013-04-22 22:28 (earthquake; M= 4.8; Tenk; dist=42.0 km) & 20.76
& 22.57 & 15.99\tabularnewline
\bottomrule
\end{longtable}

\subsection{Seismic hazard in the Mátra
Mountains}\label{seismic-hazard-in-the-matra-mountains}

The SHARE (Seismic Hazard Harmonization in Europe) was a large-scale collaborative project under the European Community's Seventh Framework Program \cite{GiaEta14a}. Its main objective was to construct a community-based seismic hazard model for the Euro-Mediterranean region. Its product, the 2013 Euro-Mediterranean Seismic Hazard Model (ESHM13) provides, among others, ground motion hazard maps for the peak ground acceleration (PGA) for different exceedance probabilities. The hazard values are referenced to a rock velocity (vs30) of 800 m/s.

For the Mátra area we downloaded the mean PGA values of the SHARE Mean Hazard Model from the online data resource \cite{GiaEta13r} for 73, 102, 475, 975 and 2475 years return periods (Table 2.). These correspond to 50\%, 39\%, 10\%, 5\% and 2\% exceedance in 50 years, respectively. 
\begin{longtable}[]{@{}rr@{}}
\caption{Average peak ground accelaration values in the Mátra Mountains
for different return periods}\tabularnewline
\toprule
Return period {[}\emph{years}{]} & PGA {[}\(m/s^2\){]}\tabularnewline
\midrule
\endfirsthead
\toprule
Return period {[}\emph{years}{]} & PGA {[}\(m/s^2\){]}\tabularnewline
\midrule
\endhead
73 & 0.016\tabularnewline
102 & 0.021\tabularnewline
475 & 0.051\tabularnewline
975 & 0.077\tabularnewline
2475 & 0.123\tabularnewline
\bottomrule
\end{longtable}

\section{Seismological measurements\label{sec:_seismo_measurements}}

Seismological data collection was performed by two Guralp CMG 3T low noise, broadband seismometers, which are sensitive to ground vibrations with flat velocity response in the frequency range 0,008-50Hz. The self noise of the seismometers is below the low noise model form 0.02Hz to 10Hz. One of these instruments (ET1H), was permanently installed in the MGGL. The installation table is fixed to the solid rock and prepared for seismological measurements (e.g. covered by a granite tile). The other instrument (hereafter GU02) was used in a measurement campaign in the first two-week of June 2017 in a measurement cabin, constructed next to the main tunnel and prepared for seismometer installation. The data collection period for ETH1 has been started on 2016-03-01 and in this paper we elaborate the data until 2017-12-15. During that period the instrument was operative for {\dnum  } days. Between 2016-09-09 - 2016-09-19 and 2016-12-10 - 2016-12-18 ET1H was out of order for 47 days. As the problems were originated in the Z direction, we have omitted the Z direction data from 2016-09-19 until 2016-11-18.

\subsection{Data analysis and site criteria}

In our analysis we use the standard seismic procedures for processing the acquired data according to \cite{BekEta15a}, e.g. the so-called Nuttal-window was applied. In this section we recall the the basic definitions. The PSD is defined as
\begin{equation}
P^{(v)}=\frac{2}{f_{s}\cdot N\cdot W}\left|V_{k}\right|^{2},
\end{equation}
where $f_{s}$ is the sampling rate, $N$ is the length of the analysed data sample, and $W=\frac{1}{N}\sum_{n=1}^{N}w[n]^{2}$ with the Nuttall window function $w[n]$. The coefficients $V_{k}=F(w[n](v[n]-\langle v\rangle)$, represent the Fourier transform $F$ of the deviation of raw data $v[n]$ from its average value $\langle v\rangle$. In our analysis PSD-s were calculated with 50s data samples according to the $f_{s}=100\,Hz$ sampling rate. We did not want to use the advantage of fast Fourier algorithm on the expense of incresing the lowest frequency. Before further processing, raw data were highpass-filtered with $f_{HP}=0.02\,Hz$. 

The Amplitude Spectral Density (ASD) can be calculated from PSD via $A^{(v)}=\sqrt{P^{(v)}}$ and ASD (PSD) can be expressed as either acceleration ($a$) or displacement ($d$) by multiplying or dividing by $\omega=2\cdot\pi\cdot f_{k}$ ($\omega^{2}$). For example, $A^{(d)}=A^{(v)}/\omega$. Therefore the mentioned ET required level, \re{eq:ETReqBF}, can be transfered easily to other spectral densities, e.g.
\begin{equation}
P^{(a)}_{BF} = 4\cdot10^{-16}\frac{m^{2}/s^{4}}{Hz}\,\,\,\text{or}\,\,\,
P^{(d)}=\left(\frac{A_{BF}}{\omega^2}\right)^2 = 4\cdot10^{-16} \omega^{-4} \frac{m^{2}}{Hz}\,\,\:\text{if}\,\,1Hz\leq f\leq10Hz\label{eq:ETReqv}
\end{equation}

\begin{figure}
\centering
\includegraphics[scale=0.5]{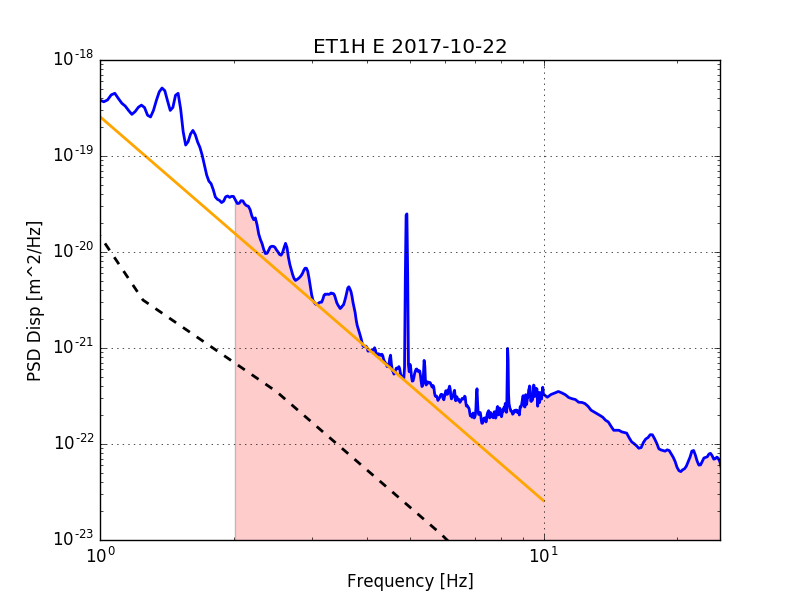}
\caption{\label{fig:rms-illusztration} Illustration of $rms$ at a displacement PSD spectrum: The blue line is the daily average of 2017-10-22 of ET1H station, E direction. The red area represents the $rms$ value from $2\,Hz$. The $rms_{2Hz}=0.209\,nm$, $rms_{2-10Hz}=0.144$, the ratio of them is $69.7\%$. The dashed line indicates the New Low Noise Model of Peterson (NLNM).}
\end{figure}

It is convenient to characterize sites in terms of $rms$ displacement, ASD acceleration and its variation. The $rms$ displacement is the square root of the integral of PSD displacement between two frequency values
\begin{equation}
rms^{(d)}=\sqrt{\frac{1}{T}\sum_{k=l}^{N/2+1}P_{k}^{(x)}},
\end{equation}
where $l$ is the cut off index, $T=\frac{N}{f_{s}}$. The usual choice is 2 Hz, as it is illustrated in (\ref{fig:rms-illusztration}).

The measurements for the ET site selection are based on the: (a) Black Forest or ET line, the lower envelope of the acceleration PSD of the Black Forest line, (\ref{eq:ETReqBF}) and (b) the corresponding $rms^{(d)}=0.1nm$ values at 2Hz:
\begin{gather*}
rms_{2Hz}^{(d)}=\sqrt{\int_{2}^{f_{s}/2}P^{(a)}_{BF}\frac{1}{\left(2\pi\right)^{4}}\frac{1}{f^{4}}df}=
\frac{A^{(a)}_{BF}}{\left(2\pi\right)^{2}}\sqrt{\int_{2}^{f_{s}/2}\frac{1}{f^{4}}df}\approx \\\frac{A^{(a)}_{BF}}{\left(2\pi\right)^{2}}\sqrt{\left[f^{-3}/(-3)\right]_{2}^{\infty}}\approx0.1nm,\label{eq:rms2hz}
\end{gather*}
where we considered that at higher frequency the displecement PSD values decrease significantly, so the change of the integration limit is a good approximation. To get a better visual picture, see Figure \ref{fig:rms-illusztration}, where it is transparent that in some cases higher frequencies can give significant contributions to the $rms$ value. 
Throughout the paper we only use the $rms$ displacement values and the upper ($d$) index is omitted in the following expressions.

\subsubsection{Modus vs. median\label{subsec:Modus-vs.-median}}

\begin{figure}
\centering
\includegraphics[scale=0.35]{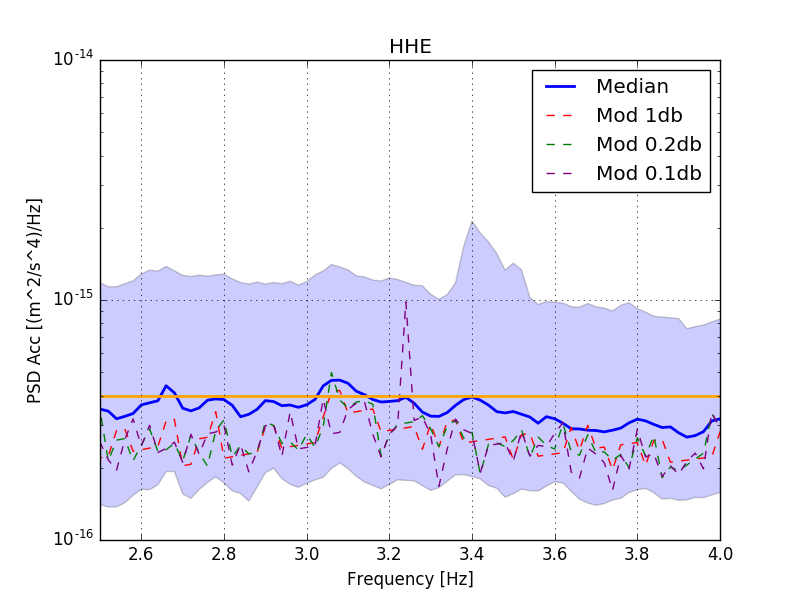}
\caption{\label{fig:MedvsMod-PSD-vel} Mode vs. Median: the modus depends on the relative bandwidth. The blue line is the median and the dashed lines are the modus  1dB, 0.2dB and 0.1dB bandwidths. The fluctuation of the modus may extend to the 10th and 90th percentiles.}
\end{figure}

In Beker et al \cite{BekEta15a} the modus of the short term averages was used for the characterisation of the typical noise level. Here we suggest to apply the median for long term purposes. In order to illustrate the differences, we use the data of the GU02 seismometer during the two-week at the 404m deep location. First the modes of two different constant relative bandwidth were calculated using half-hour averaged PSD values. This is shown together with the 10th, 50th  and 90th percentiles of the half-hour averages on Figure \ref{fig:MedvsMod-PSD-vel} between 2.5Hz and 4 Hz. One can see that the use of different relative bandwidth can modify the mode. The corresponding $rms_{2Hz}$ values can be found in Table \ref{tab:ModvsMed-rms}. 

\begin{table}
\centering
\begin{tabular}{|c|c|c|c|}
\hline 
$rms_{2Hz}\,[nm]$ & E & N & Z\tabularnewline
\hline 
\hline 
Median & 0.101 & 0.100 & 0.089\tabularnewline
\hline 
Modus $1\,dB$ & 0.088 & 0.088 & 0.074\tabularnewline
\hline 
Modus $0.2\,dB$ & 0.091 & 0.092 & 0.073\tabularnewline
\hline 
Modus $0.1\,dB$ & 0.089 & 0.092 & 0.075\tabularnewline
\hline 
\end{tabular}
\caption{\label{tab:ModvsMed-rms} The $rms$ moderately depends on the bandwidth. }
\end{table}

\begin{figure}
\centering
\includegraphics[scale=0.4]{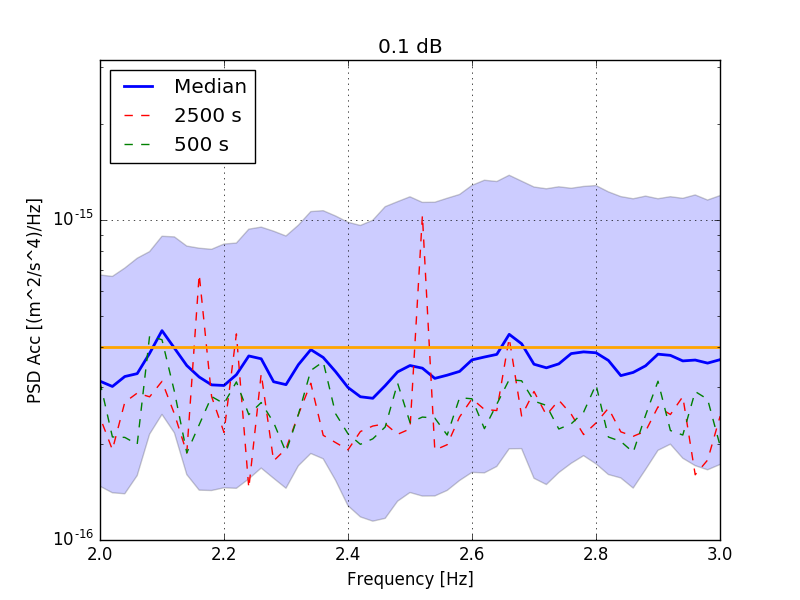}
\caption{\label{fig:ModvsMed-PSD-time} Mode vs. Median: different averaging
length. The blue line is the median and the dased lines are the modus  with 500s and 2500s averaging lengths. The fluctuation of the modus may extend to the 10th and 90th percentiles.}
\end{figure}

The dependence of the averaging time is illustrated in Figure \ref{fig:ModvsMed-PSD-time}
and the $rms_{2Hz}$ values in Table \ref{tab:ModvsMed-rms-db}, too.

\begin{table}
\centering
\begin{tabular}{|c|c|c|c|c|c|}
\hline 
$rms_{2Hz}\,[nm]$ & 500s & 1000s & 1500s & 2000s & 2500s\tabularnewline
\hline 
\hline 
0.1 dB & 0.087 & 0.088 & 0.086 & 0.087 & 0.088\tabularnewline
\hline 
0.5 dB & 0.087 & 0.086 & 0.085 & 0.086 & 0.089\tabularnewline
\hline 
\end{tabular}
\caption{\label{tab:ModvsMed-rms-db} $rms$ values for different averaging
length: The $rms$ calculated from the median is $0.101\,nm$ in the E direction.}
\end{table}

As it can be seen there is no significant difference between the calculated $rms$ values either using various relative bandwidth or averaging times. But there are some frequency values where the modes significantly differ from each other. Recalling that there is no well-defined control about the modes, namely usually just a few percent  of the PSD values could define the actual mode, we suggest to use the median instead of modes to get a better measurement for any candidate.\footnote{For a two-week interval about 700 samples of half-hour PSD-s are calculated. Considering a bandwidth with 0.1 dB, and the difference between the 10th, 90th percentiles is 1dB, we obtain 100 bins for calculating the mode. With a sufficiently uniform noise strength distribution at a given frequency 8 PSD values can define the mode. In case of several peaks with varying strength the mode can fluctuate violently. The median characterize the best/worst 50 \% of the data.} 
Futhermore, as we have seen in this section, there could be a good and easy way to calculate the medians of longer time period.

\section{Long term seismic data of MGGL\label{sec:Long-term-seismic-data}}

For our long term noise analysis, based on data for almost two years, some specific aspects should be considered. There is an ongoing re-cultivation activity in the Gyöngyösoroszi mine. As a result, the investigation and identification of various noise sources originating from human activity, machine noise, construction works, train noise, etc. proved to be difficult. In November 2016 a three-shift operation period has been started in the mine with increased human noise present also during night periods. Because of these kind of activities in our analysis we have defined three periods for each day: (a) the whole day, (b) night period (20:00 - 2:00 UTC) and (c) working period (9:00 - 15:00 UTC).

As we have already mentioned, there are two important aspects when long term characteristics of low frequency noise has to be analysed: the filtering of very noisy periods and the need of intermediate averaging over the basic averaging periods, determined by the data collection frequency of the seismometers (50s in our case and 128s for the Trillium 240 instruments used in the previous study). These questions are relevant both for the spectral and for the cumulative characteristics. The calculation of the weighted averages is sensitive to the high peaks, that can be orders of magnitudes higher that the average noise level, Considering the huge amount of data collected so far it is convenient to use intermediate averaging. Simple averaging without filtering is not suitable for spectral noise characterisation: short but noisy periods are characteristic in seismology and they can dominate the spectrum. However, an ad hoc removal of the highest peaks biases the data and is sensitive to the particular spectral distribution and to the characteristics of the site itself. Therefore it is preferable to apply a filtering method that is based on the  intrinsic properties of the data. {\it For that purpose we suggest to use the percentiles.} It is remarkable that for short term data with only one or two exceptional events, the different filtering methods lead to  similar results as we will see in section \ref{subsec:2-weeks-results}, Figure \ref{fig:PSD-2-week-night_avcomp}, where various averages of data is compared for a two-week period. 

In this section we summarize the results of our analysis of long term seismic data. Our filtering applies percentiles both for the characterisation of the typical long term properties and for the characterisation  of deviation from typical, including the spectral properties and the {\it rms} values. Results are presented in the form of the application of three types of averaging methods.  Here we have chosen the daily averages for the evaluation of long term characteristics, instead the 14x128s length averaging of Beker {\em et al}. This choice considers the natural periodicity of the data, other choices could be more suitable for operational performance of ET. 

The yearly PSD averages and also the corresponding {\em rms }values are calculated either by 
\begin{itemize}
\item averaged daily percentiles,
\item the percentiles of the daily averages, or using
\item the percentiles of the daily percentiles.
\end{itemize}
In the following we will see the consequences of the various choices. 

\subsection{Yearly changes\label{subsec:Yearly-changes}}

During the calculation of the PSD for the whole observation period of {\dnum } days raw data is kept in its original form without any kind of filtering method. In the following, two different types of PSD values are determined: (a) we calculate the percentiles of daily weighted averages from almost 1800 of 50s long PSDs to get the yearly average and the 10th, 50th and 90th percentiles; and (b) we calculate the weighted averages from daily 10th, 50th and 90th percentiles. The weight factor is the number of daily 50s long PSDs. These two kinds of calculations are different, especially when the $rms$ were calculated as a result of the fact that one high 50s piece of data can dominate the average. This will be demonstrated in this section. To avoid any misunderstanding, we indicate which averaging method was used to get PSD and $rms$ values. The difference between the evaluation methods might not show up during the analysis of shorter time periods, e.g. few days.

\subsubsection{Whole-day results\label{subsec:Yearly-Whole-day-results}}

\begin{figure}
\centering
\includegraphics[width=.45\textwidth]{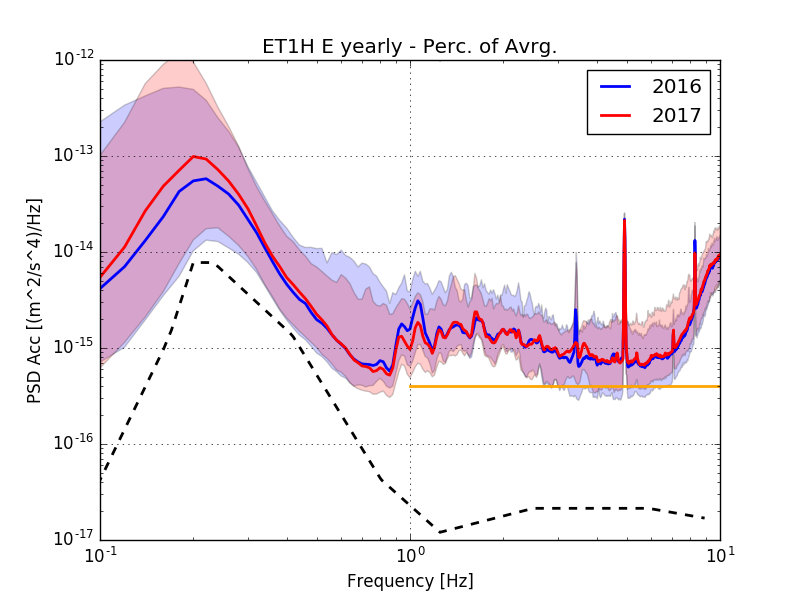}
\includegraphics[width=.45\textwidth]{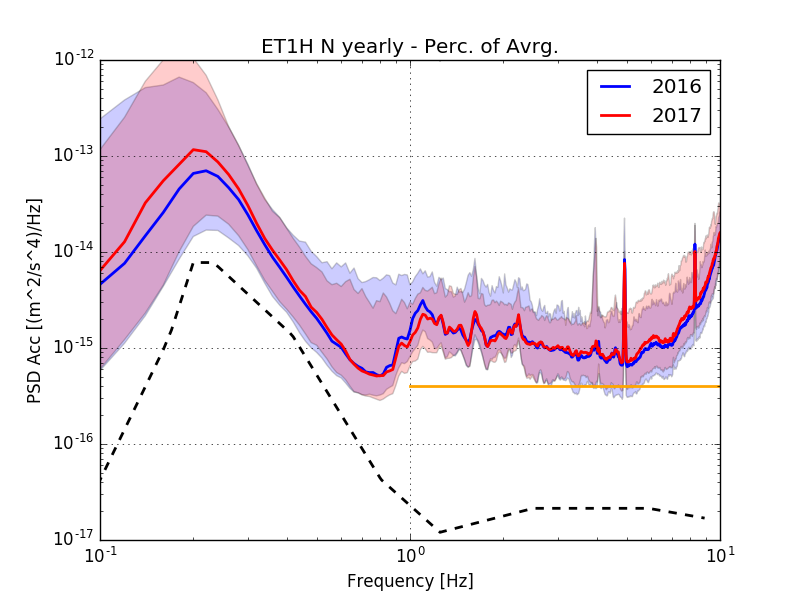}

\includegraphics[width=.45\textwidth]{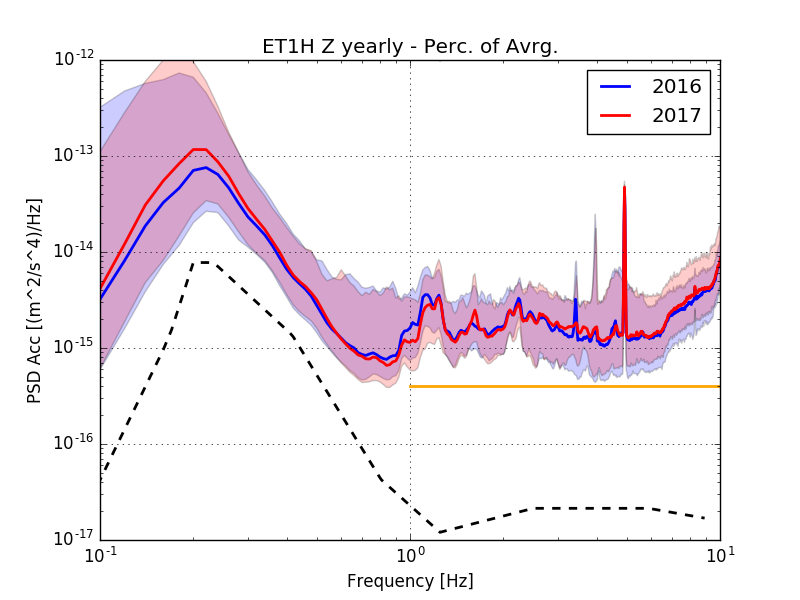}

\includegraphics[width=.45\textwidth]{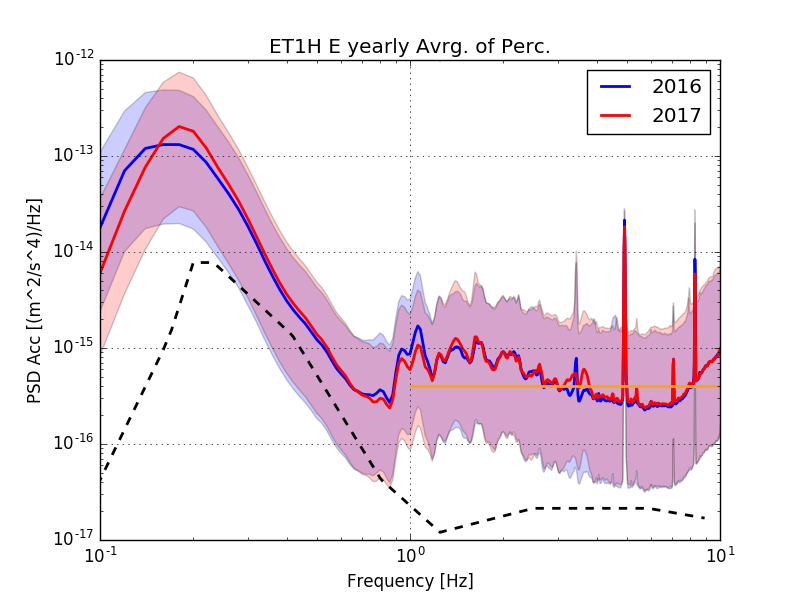}
\includegraphics[width=.45\textwidth]{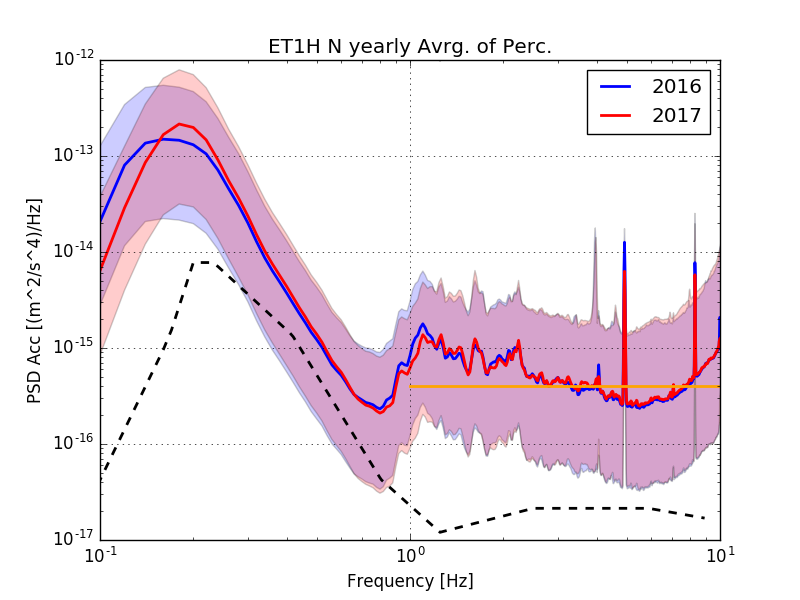}

\includegraphics[width=.45\textwidth]{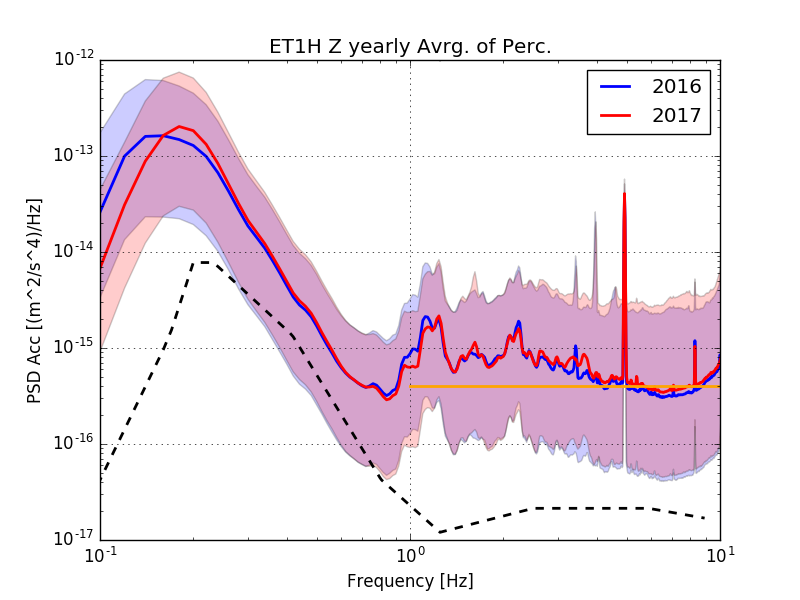}
\caption{\label{fig:Yearly_PSD} The first three figures were calculated as the 10th, 50th and 90th percentiles of the daily averages, in the last three ones are the averages of the daily 10th, 50th and 90th percentiles in the East, North and vertical Z directions, respectively. The PSD-s calculated by the second method are closer to the required  Black Forest line.}
\end{figure}

The acceleration PSDs for the two-year observation period are shown in Fig. \ref{fig:Yearly_PSD}. The corresponding $rms_{2Hz}$ values can be found in Table \ref{tab:Yearly-rms}, where also the median of the daily medians is indicated. The results differ significantly for the calculation methods introduced above. 

The acceleration PSD-s calculated by the second type of averaging method are closer to the required  Black Forest line, even with the 88m deep laboratory. Also remarkable, that the deviation from the median, indicated by the lines of the 10th and 90th percentiles, is significantly larger, too. The broader area between the 10\% and 90\% deviation curves characterises the daily variation during a year, while the slimmer stripes of the first three figures are characterising the yearly variation of the daily averages during the year. None of the figures indicate a significant difference between the years in the average noise level: thick blue and red curves in the middle of the stripes for 2016 and 2017 are close to each other. This is in spite of the increased activity in the mine (three shifts with doubled crew in 2017 instead of zero or one shifts during 2016). The effect of work-night and seasonal changes will be investigated in the following subsection. 

\begin{figure}
\centering
\includegraphics[scale=0.5]{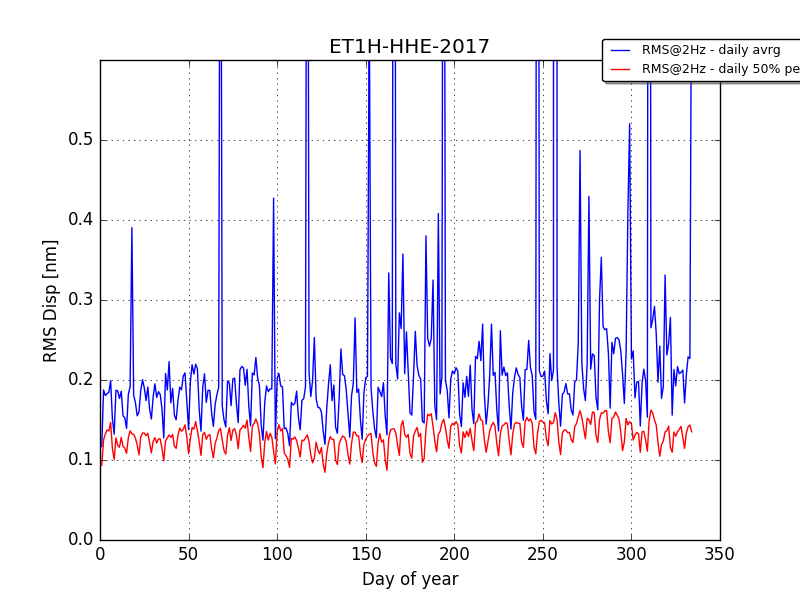}
\caption{\label{fig:rms-2017-E}$rms_{2Hz}$ from daily average (blue) and daily 50th percentile (red) data. If we ignore the 13 highest differences from the whole 334 days the average ratio between the two $rms$-s is 1.51 (with standard deviation: 0.26). This is close to the ratio of the $rms$ values from the second and third column of Table \ref{tab:Yearly-rms}, which is between 1.45-1.53.}
\end{figure}

As it can be seen in Table \ref{tab:Yearly-rms}, the three types of $rms$ values are very different, the highest one was measured in 2017, E-direction. To illustrate the reason of this difference with the simple averaging, we plot the daily $rms$ of 2017 in Figure \ref{fig:rms-2017-E}. The protrusions are in the order of 100 nm and can modify the yearly average and are due to short periods of increased seismic activity. As one can see in the figures, there is a remarkable difference between the $rms$ from daily averages and $rms$ from median of days so we have to specify exactly the calculation method. This value could also characterize a site, because it is sensitive to protrusions in daily averages.

Let us recall, that the $rms_{2Hz}$ value of horizontal noise the in this depth was $0.12 nm$ in the evaluation of the former, short term measurements  \cite{Bek13t,BekEta15a,ETdes11r}. However, that value was calculated  form the mode of the data! We have seen in section \ref{subsec:Modus-vs.-median} that difference of mode and median is related to differences in noise strength distribution and we will see in section \ref{sec:Conclusion1} that in this case mode gives lower values. One should use more indicators for a proper characterisation and choose the best for the ET purposes, including the comparison of different sites. 

\begin{table}
\centering
\begin{tabular}{|c|c|c|}
\hline 
Yearly & 2016 & 2017\tabularnewline
\hline 
E & 0.31, 0.187, 0.129 & 22.4, 0.187, 0.13\tabularnewline
\hline 
N & 0.36, 0.191, 0.129 & 14.96, 0.195, 0.128\tabularnewline
\hline 
Z & 0.38, 0.248, 0.175 & 0.83, 0.245, 0.162\tabularnewline
\hline 
\end{tabular}

\begin{tabular}{|c|c|c|}
\hline 
Night & 2016 & 2017\tabularnewline
\hline 
E & 0.243, 0.129, 0.129 & 0.153, 0.161, 0.13\tabularnewline
\hline 
N & 0.307, 0.128, 0.129 & 0.174, 0.134, 0.128\tabularnewline
\hline 
Z & 0.349, 0.248, 0.175 & 0.179, 0.245, 0.162\tabularnewline
\hline 
\end{tabular}

\begin{tabular}{|c|c|c|}
\hline 
Working & 2016 & 2017\tabularnewline
\hline 
E & 0.377, 0.22, 0.164 & 0.485, 0.225, 0.169\tabularnewline
\hline 
N & 0.391,0.227, 0.169 & 0.398, 0.233, 0.171\tabularnewline
\hline 
Z & 0.369, 0.236, 0.225 & 0.596, 0.299, 0.220\tabularnewline
\hline 
\end{tabular}
\caption{\label{tab:Yearly-rms}The $rms_{2Hz} [nm]$ values from the yearly, yearly night and yearly working averages, using three types of calculation method: first $rms$ calculated from the average of the daily average; second calculated from the 50th percentile of daily average; and third $rms$ calculated from the average of daily 50th percentile. The ET requirement 0.1 nm was calculated by integrating the mode of the data (see section \ref{sec:Conclusion1}). }
\end{table}

\subsubsection{Night and working results\label{subsec:Yearly-Night-and-working}}

In the work period (9:00 - 15:00 UTC) one ought to have the highest level of civilization noises and the night (20:00 - 2:00 UTC) might have the lowest (recalling the three-shift work in 2017). Otherwise we may expect that night periods provide a good approximation of ``noiseless'' time, because of the reduction of external and internal noise sources. 

\begin{figure}
\centering
\includegraphics[scale=0.25]{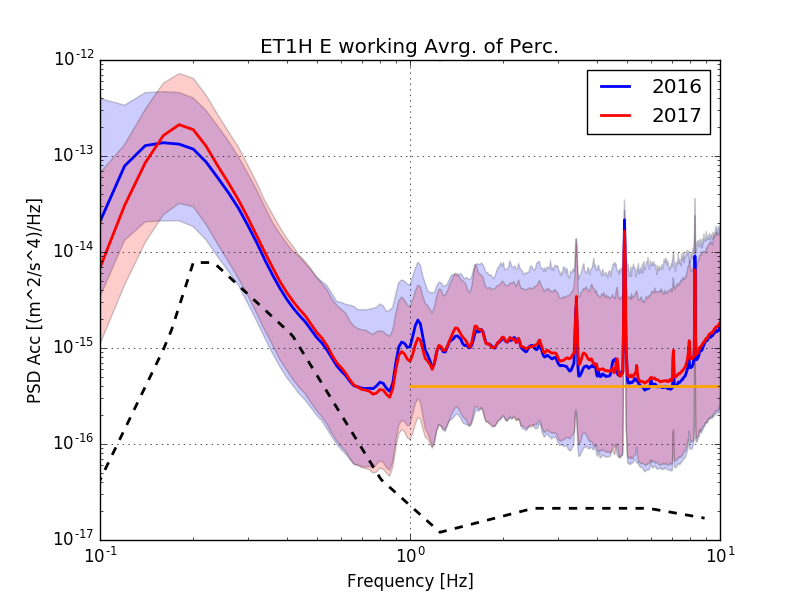}
\includegraphics[scale=0.25]{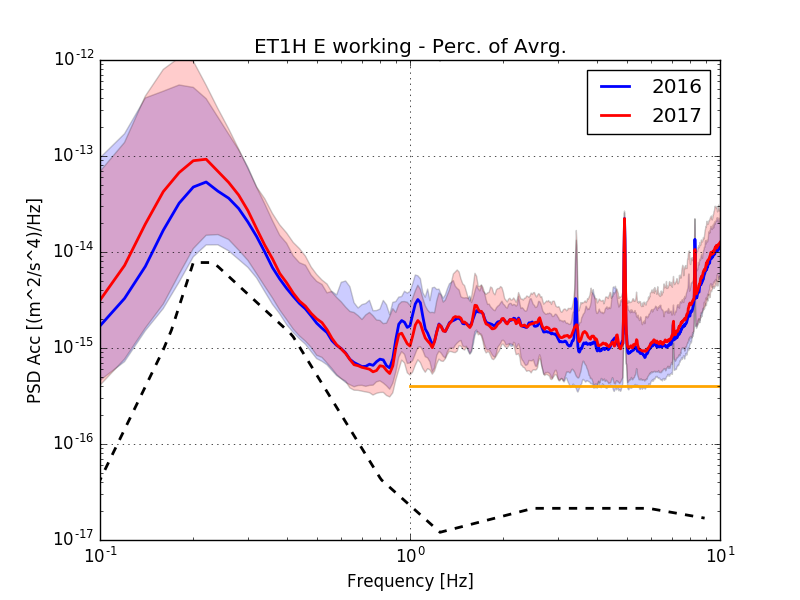}

\includegraphics[scale=0.25]{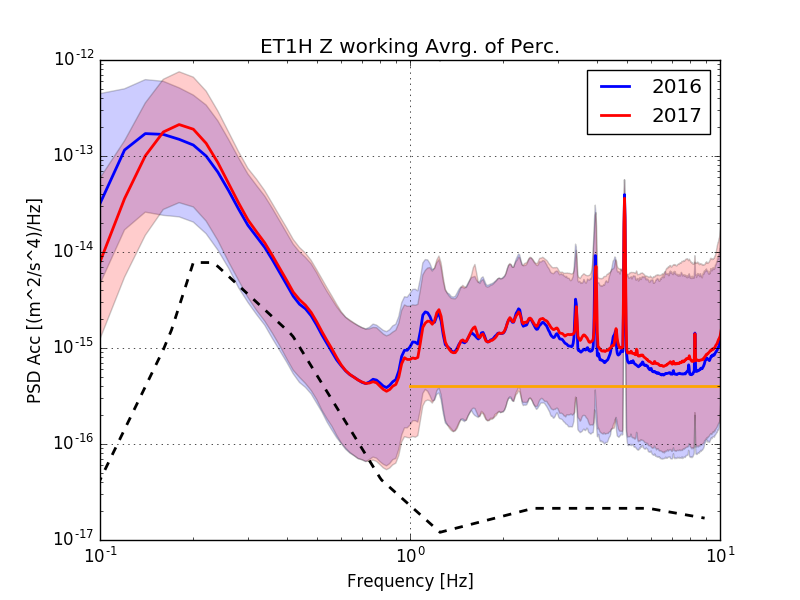}
\includegraphics[scale=0.25]{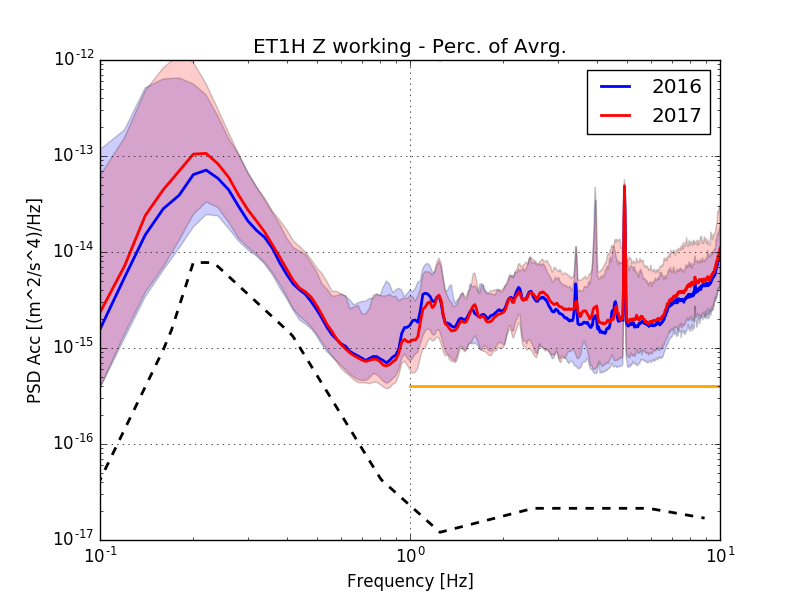}

\includegraphics[scale=0.25]{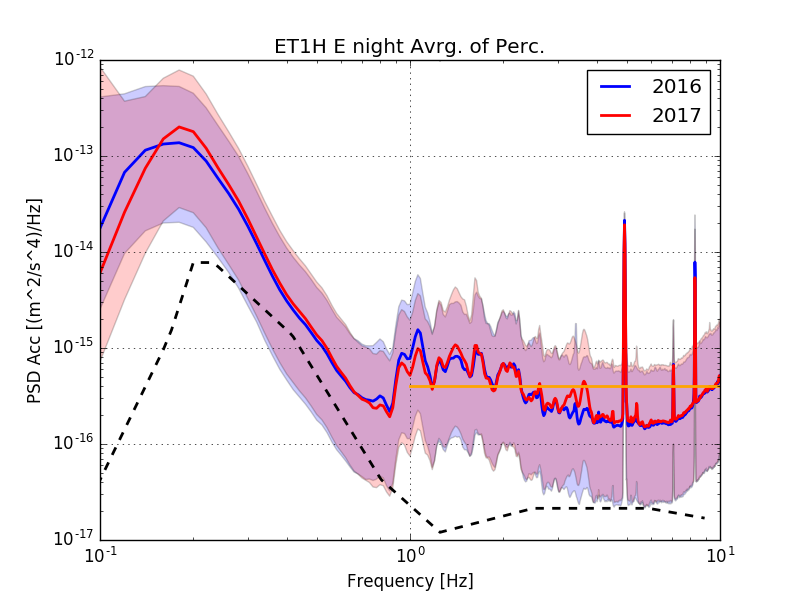}
\includegraphics[scale=0.25]{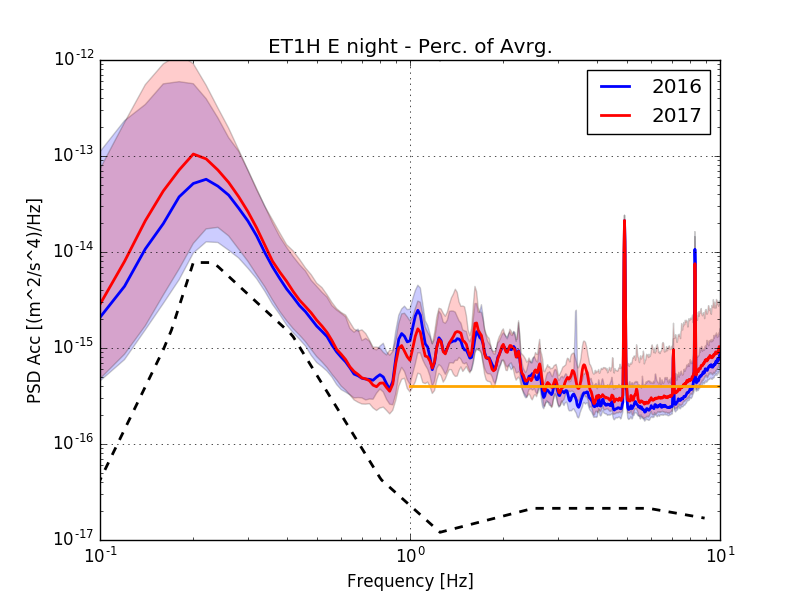}

\includegraphics[scale=0.25]{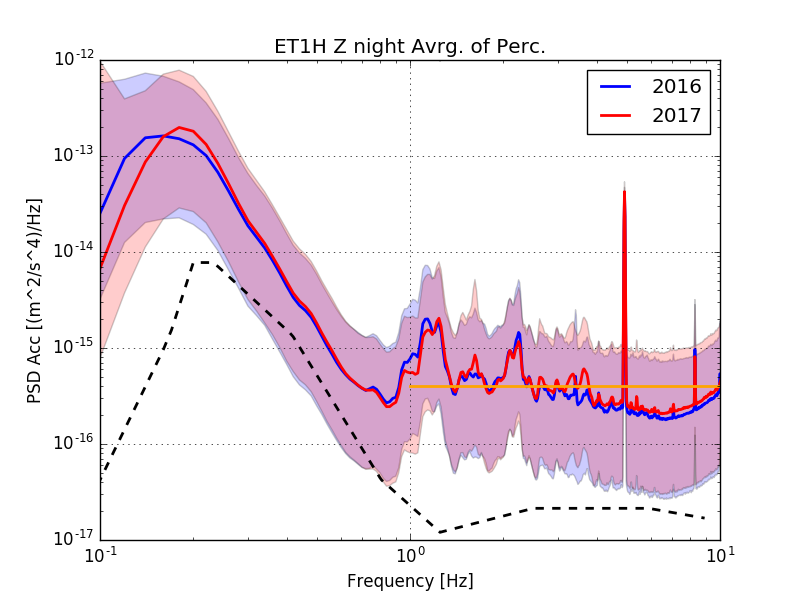}
\includegraphics[scale=0.25]{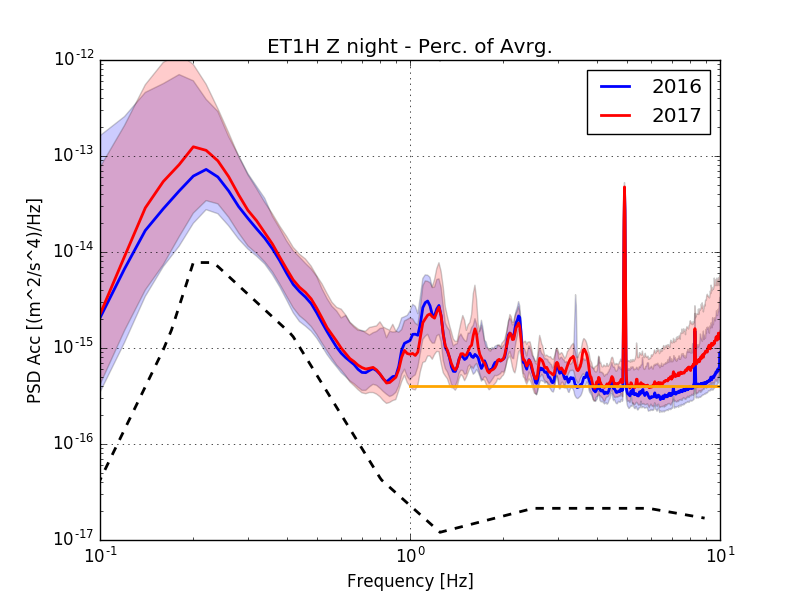}
\caption{\label{fig:Yearly-work-night-PSD}Acceleration PSD of working and night periods: The plots at the right hand side are percentiles of the daily averages and at the left hand side are calculated as the average of daily percentiles. Eastern and Z directions are shown.}
\end{figure}

The acceleration PSD of the work and night periods for 2016 and 2017 are shown in Figure \ref{fig:Yearly-work-night-PSD}. Here on the left and right sides are the averaged daily percentiles and the percentiles of the daily averages, respectively. The first four figures show the spectrum of the working periods for one of the horizontal directions and also for the Z direction. The increased human activity in 2017 is more apparent from the median of daily averages on the right hand side. In the Z direction data the problematic period of 2016 autumn, mentioned in section \ref{sec:_seismo_measurements}, is missing.

At night the typical MGGL noise fulfils the ET requirement between 2-8 Hz with the average of medians. This is the seismometers data from depth of 88m, therefore these results should be properly interpreted, not forgetting the differences of median and mode. We may expect that the mode of night periods is lower, and will not reach the Black Forest in the whole 1-10Hz intervall.

\begin{figure}
\centering
\includegraphics[width=0.45\textwidth]{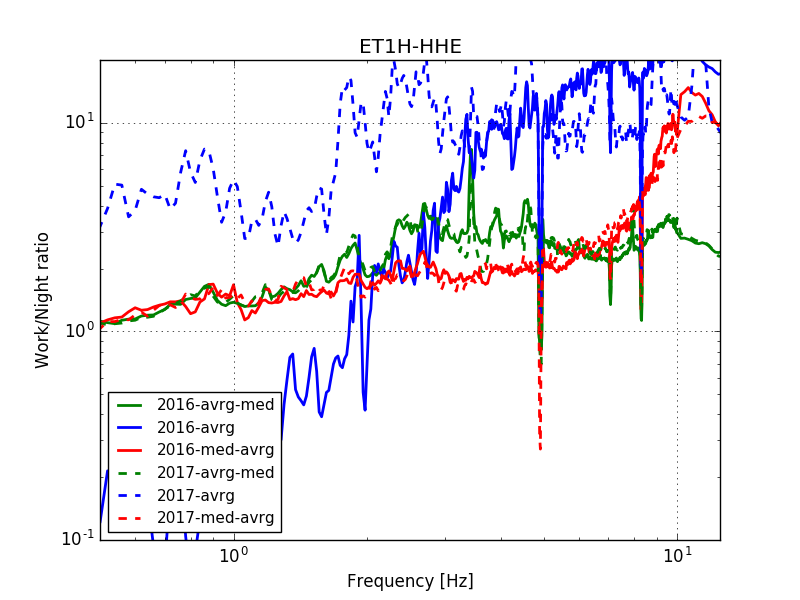}
\includegraphics[width=0.45\textwidth]{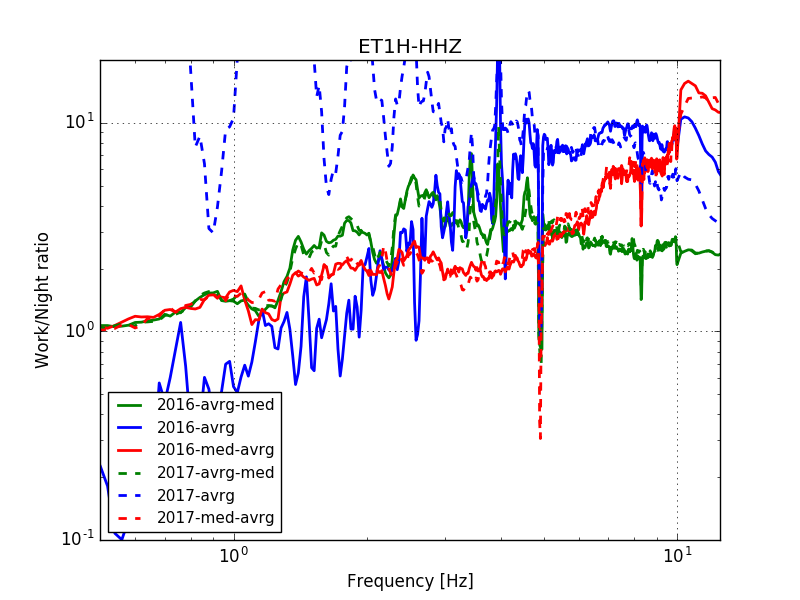}
\caption{\label{fig:Yearly-work-night-ratio}Acceleration PSD ratio of night and working: in 2016 (solid lines) and 2017 (dashed lines). The blue lines are calculated from the average of daily average, the greens from the medians of daily averages and the red ones from the averages of daily medians.}
\end{figure}

The relation of the human or civilization noises to the noises of other origin is best seen on the ratio of night and work periods, as it is shown on Figure \ref{fig:Yearly-work-night-ratio}. The average of the daily medians (green) and median of the daily averages (red) are different from each other in the different regions, but close to each other in the two years. However, when the average of the daily averages is considered (blue) then there is a big difference between the two years. The ratio based on the averages of daily medians (red) shows the increasing noise ratio at higher frequencies, therefore it seems to be most informative for our purpose, indicating the human activity. The same plot can be seen in the next section, when two-week of data is compared at different depths of the mine on Figure \ref{fig:2-week-PSD-night-work-ratio}. It is remarkable, that  there is no uniform indication of the increased human activity in 2017 on these figures.

\subsubsection{Seasonal changes \label{subsec:Yearly-season-change}}

With almost two-year of collected data seasonal changes can be studied, too. We note that only one winter period is present, because the data collection has been started in March 2016. There was no overlap in the two periods so the one winter period is from 2016-12-23 to 2017-03-20 (90 days). Furthermore, the Z directional data from observation run in autumn 2016 should be treated with caution, because of the temporary malfunction of ET1H. 

\begin{table}
\centering
\begin{tabular}{|c|c|c|c|c|}
\hline 
 & Spring & Summer & Autumn & Winter\tabularnewline
\hline 
2016 & 92 & 84 & 57(38) & 29 \tabularnewline
\hline 
2017 & 92 & 94 & 68 & 79 \tabularnewline
\hline 
\end{tabular}
\caption{\label{tab:Season-days-num}Number of the observation days in the various seasons, except the winter period. We calculated the winter period from 2016-12-23 to 2017-03-20 which was 90 days.}
\end{table}

\begin{figure}
\centering
\includegraphics[scale=0.25]{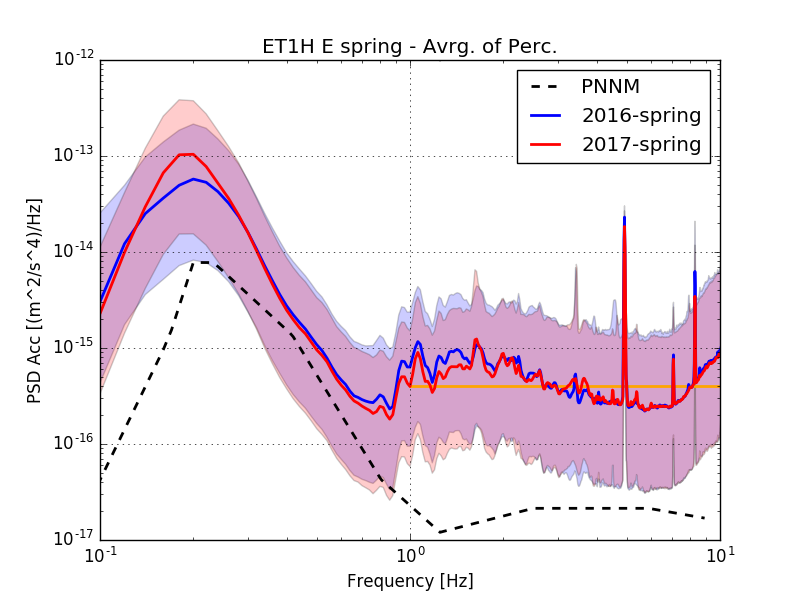}
\includegraphics[scale=0.25]{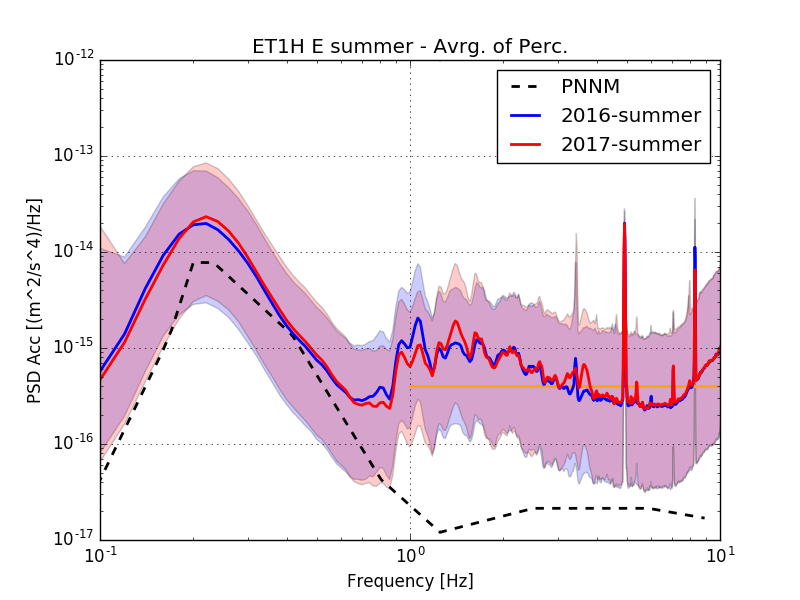}

\includegraphics[scale=0.25]{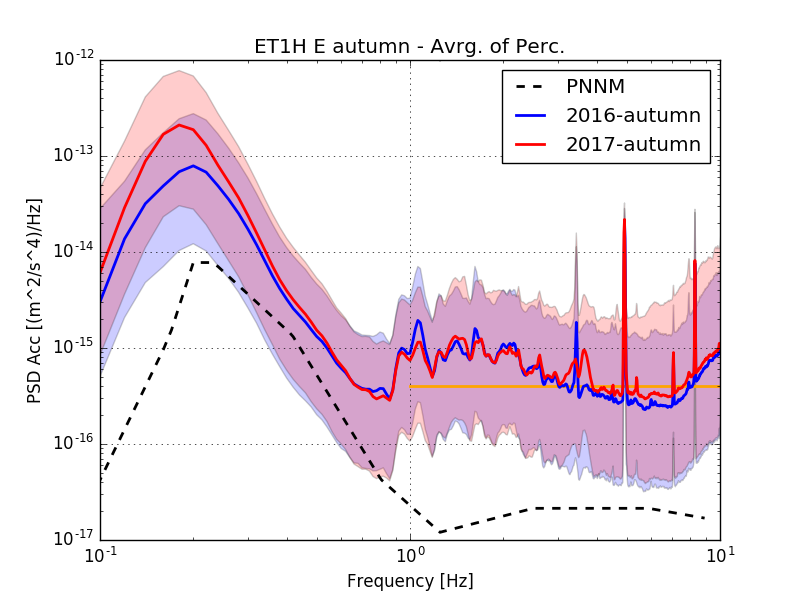}
\includegraphics[scale=0.25]{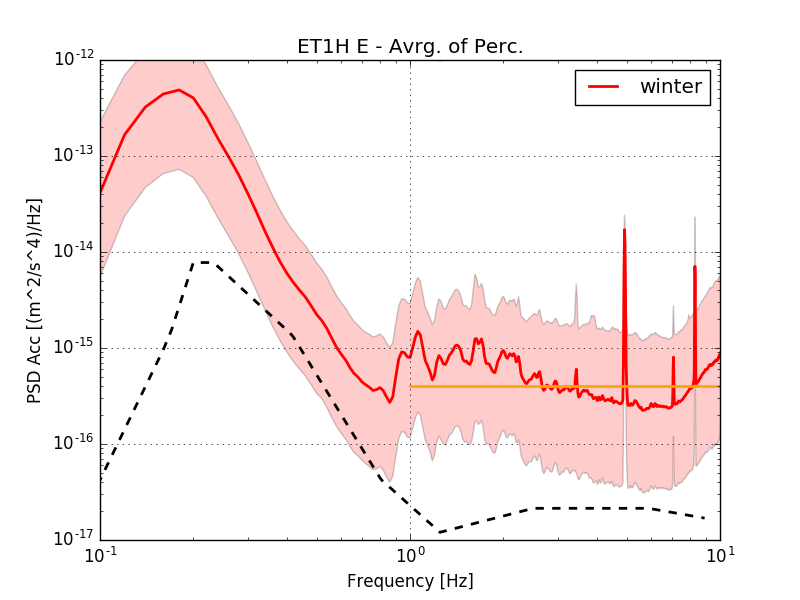}
\caption{\label{fig:Season-PSD}The East acceleration PSD of the different seasons. The averages of the 10th, 50th and the 90th percentiles are shown. The winter plot is for the single 2017 data.}
\end{figure}

For the other seasons, there is enough overlapping to be able to see any yearly changes and to compare the seasonal differences. The PSD values in one of the horizontal directions are shown in Fig. \ref{fig:Season-PSD} and the $rms$ values in Table \ref{tab:season-rms}. 

According to the PSD figures one can say that there are no significant seasonal changes in the studied frequency interval, only in higher (>10Hz) frequencies were visible differences. It is remarkable, that in the Mátra region the average wind speed is moderate and does not show seasonal differences. It is also important, that there are no leaves on most of the trees in the winter period in Mátra. 

\begin{table}
\centering
\begin{tabular}{|c|c|c|c||c||c|}
\hline 
Year & Season & PoD & E-direction & N-direction & Z-direction\tabularnewline
\hline 
\hline 
2016 & Spring & Daily & 0.367, 0.184, 0.125  & 0.422, 0.191, 0.127 & 0.315, 0.24, 0.156\tabularnewline
\cline{3-6} 
 &  & Night & 0.128, 0.122, 0.101  & 0.156, 0.123, 0.101 & 0.156, 0.149, 0.122\tabularnewline
\cline{3-6} 
 &  & Work & 1.609, 0.221, 0.163  & 1.83, 0.233, 0.173 & 1.363, 0.296, 0.215\tabularnewline
\cline{2-6} 
 & Summer & Daily & 0.322, 0.194, 0.133  & 0.368, 0.198, 0.133 & 0.353, 0.256, 0.17\tabularnewline
\cline{3-6} 
 &  & Night & 2.462, 0.131, 0.108  & 3.837, 0.129, 0.105  & 4.383, 0.162, 0.134\tabularnewline
\cline{3-6} 
 &  & Work & 0.603, 0.23, 0.167  & 0.6, 0.244, 0.17  & 0.415, 0.316, 0.22\tabularnewline
\cline{2-6} 
 & Autumn & Daily & 0.388, 0.199, 0.138 & 0.491, 0.201, 0.136 & 0.757, 0.309, 0.257\tabularnewline
\cline{3-6} 
 &  & Night & 3.075, 0.144, 0.115  & 4.04, 0.141, 0.111 & 4.232, 0.252, 0.236\tabularnewline
\cline{3-6} 
 &  & Work & 1.558, 0.23, 0.173 &  2.23, 0.239, 0.175 &  0.889, 0.341, 0.288\tabularnewline
\hline 
\hline 
2017 & Spring & Daily & 0.306, 0.181, 0.122  & 0.332, 0.19, 0.119  & 0.267, 0.227, 0.148\tabularnewline
\cline{2-6} 
 &  & Night & 0.262, 0.137, 0.1  & 0.579, 0.139, 0.095  & 0.242, 0.163, 0.116\tabularnewline
\cline{2-6} 
 &  & Work & 1.92, 0.223, 0.155  & 1.426, 0.249, 0.157  & 0.918, 0.29, 0.198\tabularnewline
\cline{2-6} 
 & Summer & Daily & 42.127, 0.199, 0.135  & 28.064, 0.205, 0.134  & 1.466, 0.261, 0.17\tabularnewline
\cline{2-6} 
 &  & Night & 0.201, 0.137, 0.11  & 0.352, 0.138, 0.107  & 0.359, 0.174, 0.134\tabularnewline
\cline{2-6} 
 &  & Work & 6.517, 0.242, 0.174  & 1.818, 0.253, 0.176  & 46.333, 0.326, 0.228\tabularnewline
\cline{2-6} 
 & Autumn & Daily & 0.38, 0.218, 0.139  & 0.331, 0.225, 0.135 & 0.492, 0.281, 0.173\tabularnewline
\cline{2-6} 
 &  & Night & 0.27, 0.15, 0.115  &  0.336, 0.15, 0.108 & 0.25, 0.179, 0.137\tabularnewline
\cline{2-6} 
 &  & Work & 3.508, 0.278, 0.191  & 2.058, 0.292, 0.19 & 3.61, 0.358, 0.249\tabularnewline
\hline 
\end{tabular}
\caption{\label{tab:season-rms} The seasonal daily/night/working $rms_{2Hz} [nm]$ values, using three calculation methods: first $rms$ calculated from the average of the daily average; second calculated from the median (50th percentile) of daily average; and third $rms$ calculated from the average of daily 50th percentile. The mode related ET requirement is $0.1 nm$. }.
\end{table}

Based on the $rms$ values we can say that the changing in yearly seasons is only based on the three-shift working, therefore the difference caused by the human activities can be evaluated. The high $rms$ values in the averages are caused due to short periods of high seismic activity. 

\subsection{Comparing the deep and shallow \label{subsec:2-weeks-results}}

\begin{figure}
\centering
\includegraphics[width=0.6\textwidth]{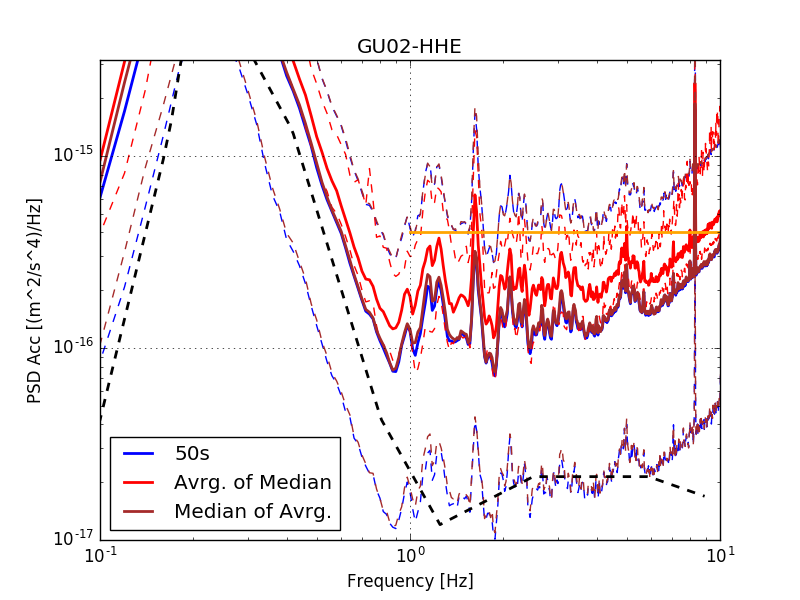}
\caption{\label{fig:PSD-2-week-night_avcomp}The differences between the averaging methods is demonstrated here: the blue line shows the percentiles from all the 50s PSD-s while the purple one is the average of the daily medians. The median of the daily averages is indicated with red. The 10th and 90th percentiles are shown with dashed lines of the same color.}
\end{figure}

\begin{figure}
\centering
\includegraphics[width=0.8\textwidth]{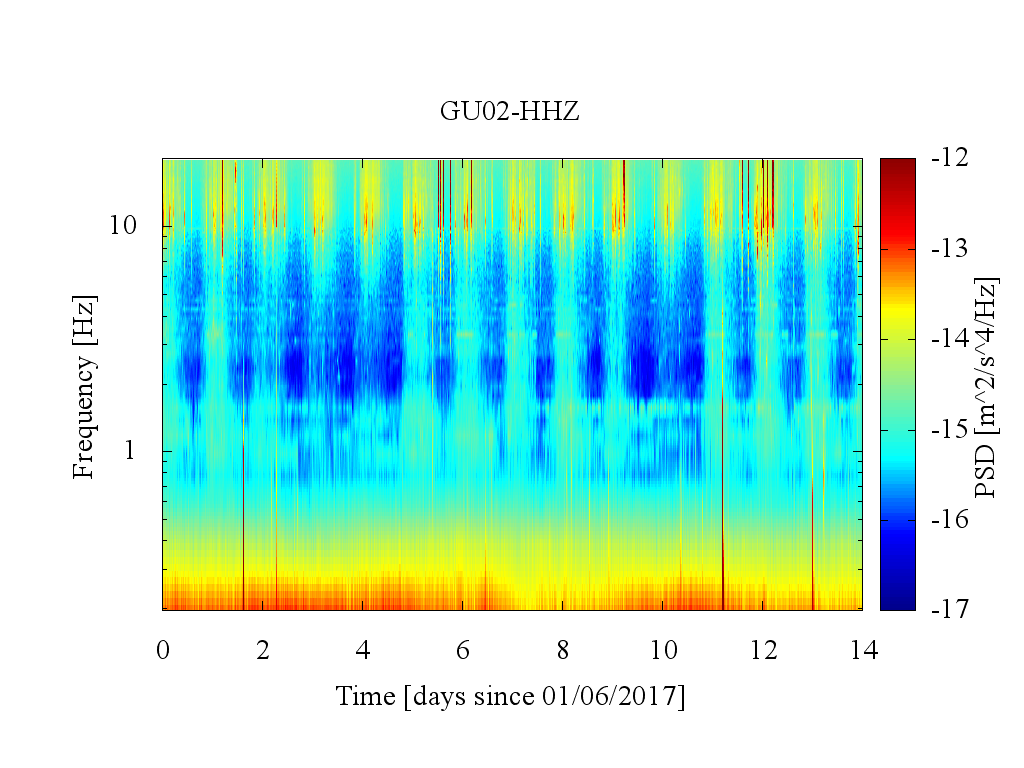}
\caption{\label{fig:2-week-contour} Two-week spectrogram of the GU02 seismometer.}
\end{figure}

In this section the GU02 (-404m) results from a two-week observation run (2017-06-01 - 2017-06-15) is compared to the parallel MGGL measurements of ET1H (-88m). The specrogram of the period is shown on Figure \ref{fig:2-week-contour}. Because of the shorter time interval, we can calculate averages and medians from shorter time averaging. In Figure \ref{fig:PSD-2-week-night_avcomp} we demonstrate the differences between the various methods comparing the percentiles of 50s periods (blue), the daily averages (red) and the average of the daily medians (red), used for long term data evaluation. One can see, that the purple and blue curves overlap, while the red one is different. 

From the whole-day PSD-s, the percentiles of the hourly averages are shown for the 404 m deep station. We can see that the ET requirement level is satisfied between 2-6 Hz for the median. The mode-median differences will be addressed in Section \ref{sec:Conclusion1}.

\begin{figure}\centering
\includegraphics[width=0.45\textwidth]{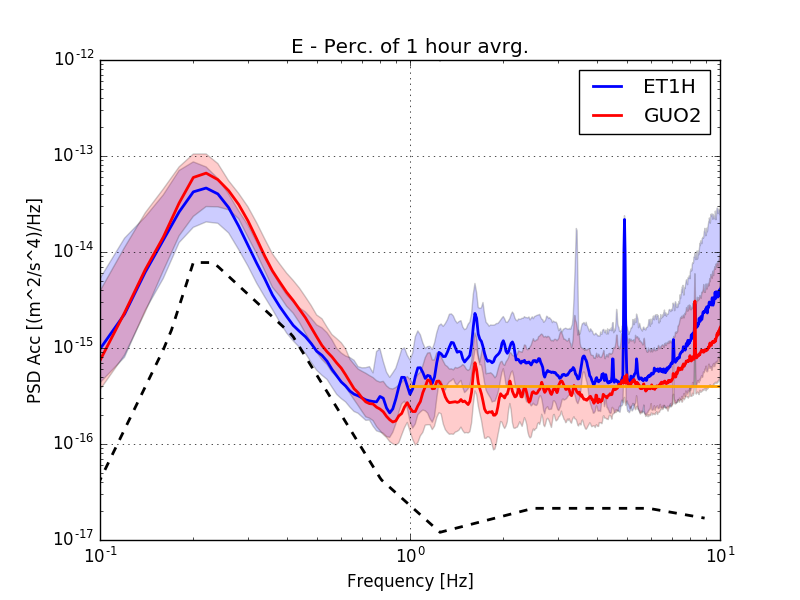}
\includegraphics[width=0.45\textwidth]{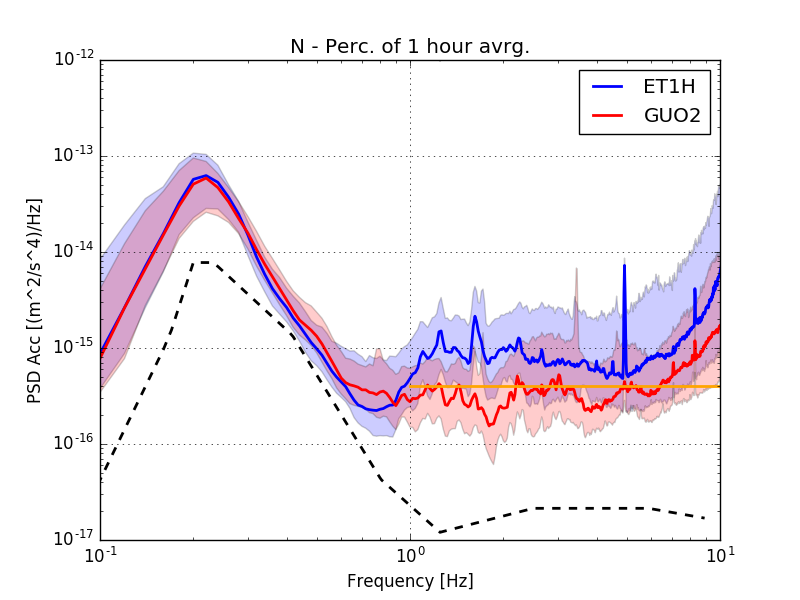}

\includegraphics[width=0.45\textwidth]{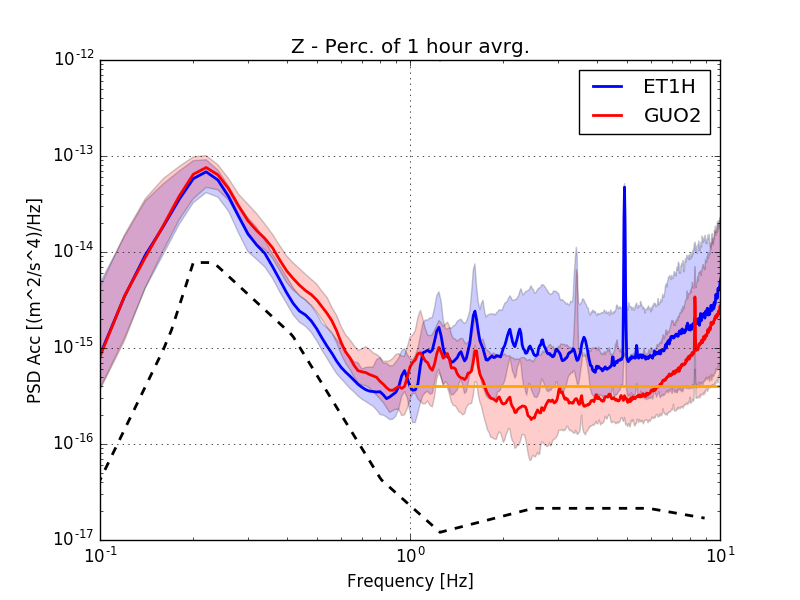}
\caption{\label{fig:2-week-PSD} Two-week PSD-s: Calculated from the 10th, 50th and 90th percentiles of one hour averages of the ET1H (88m) and GU02 (400m) seismometers. The other averaging methods were omitted here.}
\end{figure}

The corresponding $rms_{2Hz}$ values are tabulated in Table \ref{tab:rms-2-week}. These are consistent with the results of the previous study in spite of the mode-median differences. It is remarkable that the Z direction is less noisy than the horizontal directions at the deeper location. 

\begin{table}
\centering
\begin{tabular}{|c|c|c|c|c|}
\hline 
Station & Direction & Daily & Night & Work\tabularnewline
\hline 
\hline 
ET1H & E & 0.151 & 0.12 & 0.203 \tabularnewline
\cline{2-5} 
 & N & 0.159 & 0.119 & 0.214\tabularnewline
\cline{2-5} 
 & Z & 0.181 & 0.138 & 0.26\tabularnewline
\hline 
\hline 
GU02 & E & 0.101 & 0.079 & 0.146\tabularnewline
\cline{2-5} 
 & N & 0.101 & 0.077 & 0.149\tabularnewline
\cline{2-5} 
 & Z & 0.09 & 0.066 & 0.137\tabularnewline
\hline 
\end{tabular}
\caption{\label{tab:rms-2-week}$rms_{2Hz} [nm]$ values of the 2 weeks data collection period 2017-06-01 - 2017-06-15. For the whole day results one hour averaging was used instead of daily and the $rms$ values are calculated from the median of the one hour periods. The night and work median were calculated from the averages of daily night and work medians.}
\end{table}

\begin{figure}
\centering
\includegraphics[width=0.45\textwidth]{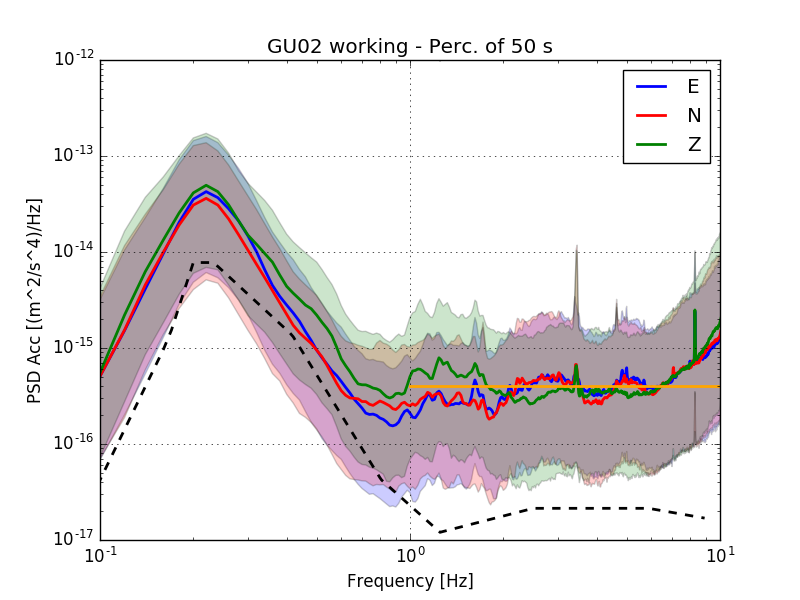}
\includegraphics[width=0.45\textwidth]{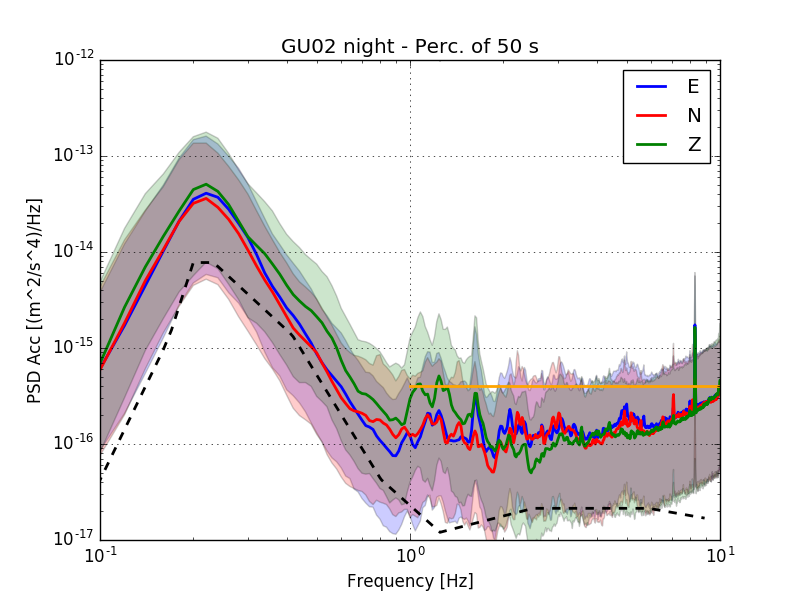}
\caption{\label{fig:2-week-PSD-night-work} PSD values of the two-week period of GU02 seismometer (-400m), from the 50s long night and work percentiles for the three directions.}
\end{figure}
\begin{figure}
\centering
\includegraphics[scale=0.25]{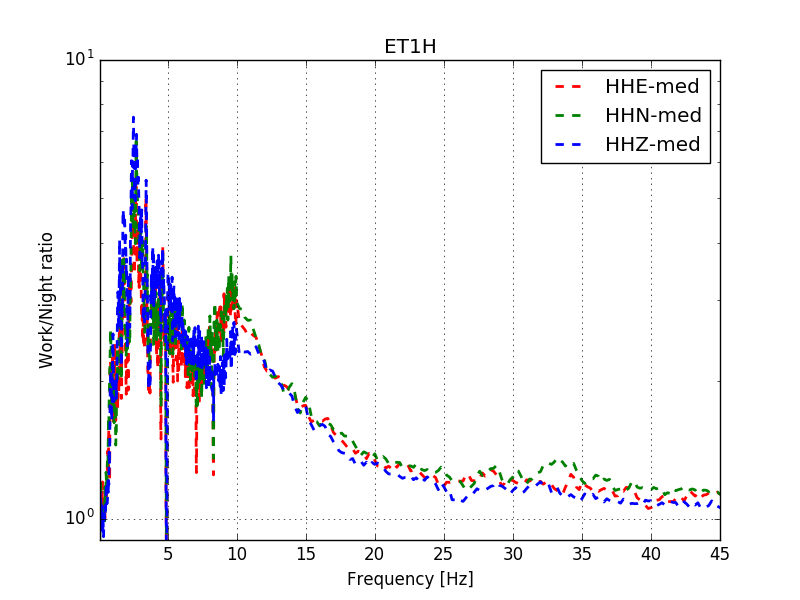}
\includegraphics[scale=0.25]{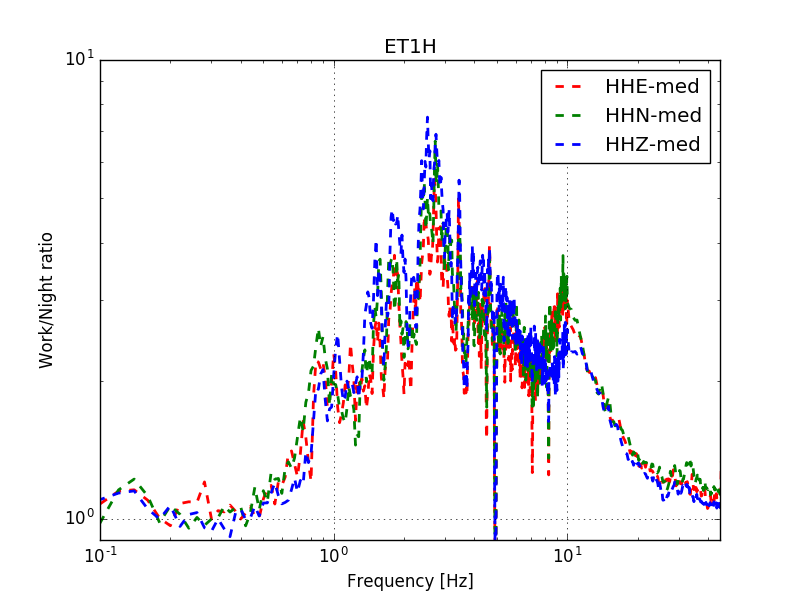}

\includegraphics[scale=0.25]{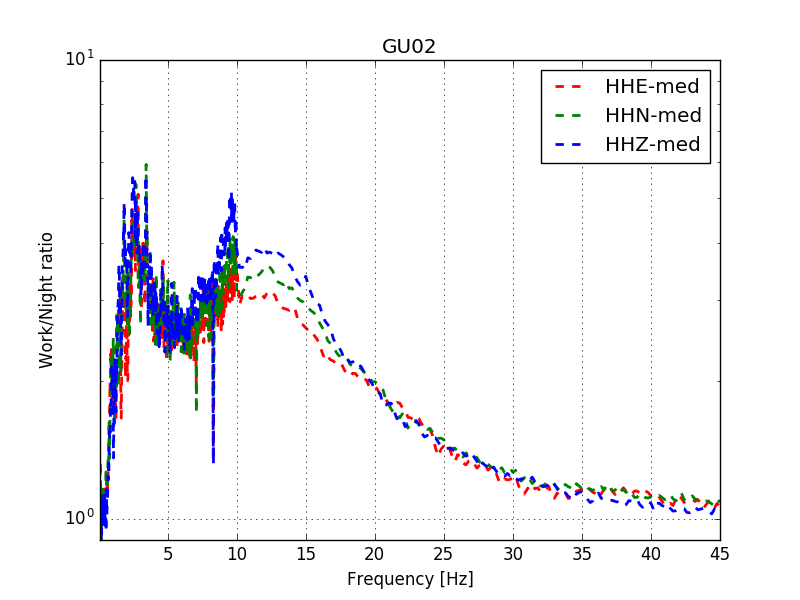}
\includegraphics[scale=0.25]{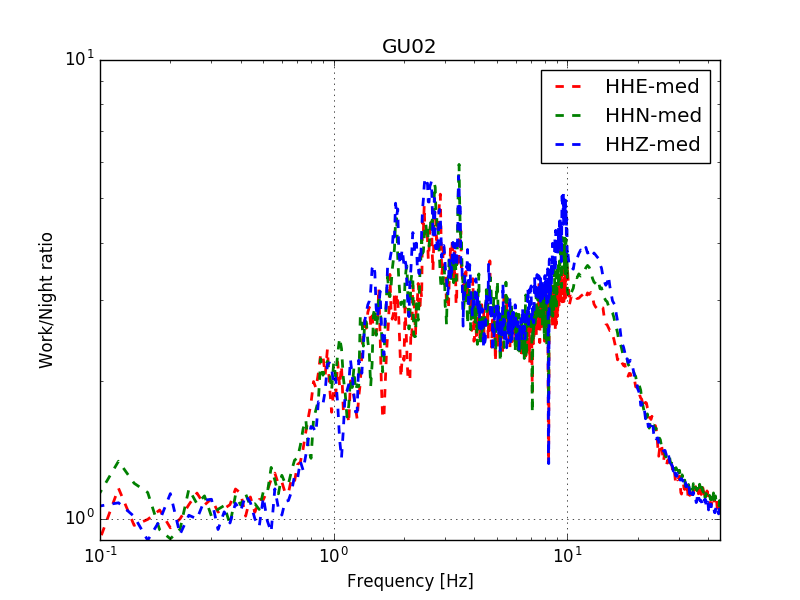}
\caption{\label{fig:2-week-PSD-night-work-ratio} Ratio of night and work periods for the two-week at ET1H (88m) and GU02 (400m) stations: The dashed lines are calculated from the median of daily night and working averages. The horizontal scales are different at the left and right hand side.}
\end{figure}

The effects of the human activity can be compared  in Figure \ref{fig:2-week-PSD-night-work}. Here the shortest,  50s long averages were used without intermediate averaging. The acceleration PSD noise ratios  of the night and work periods in the different directions  are shown in Figure \ref{fig:2-week-PSD-night-work-ratio}. 

Because of the low number of days we also calculated the working and night results with the 50s time intervals in the three directions in Figure \ref{fig:2-week-PSD-night-work}{fig:2-week-PSD-night-work}. The horizontal components have the same characteristics but the vertical one differs between 1 Hz and 3 Hz. We also can see that these results consistent with Figure \ref{fig:2-week-PSD}. 

At this point the $rms_{2Hz}$ is calculated for the 10th, 50th and 90th percentiles of the night period given in Table \ref{tab: rms-night-2-week}. The value of the upper envelope, the 90th percentile, indicates that in 90\% of 50s intervals during the two-weeks are below this value. E.g in the se direction it is below the required ET limit which is a significant information regarding the operation conditions of ET.

\begin{table}
\centering
\begin{tabular}{|c|c|c|c|}
\hline 
Direction & 10th & 50th & 90th\tabularnewline
\hline 
\hline 
E & 0.026 nm & 0.064 nm & 0.121 nm\tabularnewline
\hline 
N & 0.025 nm  & 0.061 nm & 0.118 nm\tabularnewline
\hline 
Z & 0.022 nm  & 0.053 nm & 0.100 nm\tabularnewline
\hline 
\end{tabular}
\caption{\label{tab: rms-night-2-week} $rms_{2Hz}$ values of the 10th, 50th, 90th percentiles of the 50 second long PSD-s for GU02 night.}
\end{table}

\section{Performance summary of the Mátra site \label{sec:Conclusion1}}

The Einstein Telescope, a planned third generation of gravitational wave observatory, is designed for underground operation in order to reduce the environmental noise at the frequency range 1-10Hz. How silent could a location be? More accurately: what is a reliable estimate of the long term noise level in this frequency range that could be achieved under the  operational conditions of the instrument? This is the basic question of the site selection. Our evaluation of long term seismological measurements in the Mátra Gravitational and Geophysical Laboratory focuses to the answer of this question. 

In this paper we have presented some general information related to the geophysical environment of the Mátra site.  The surface lithology is hard rock, vulcanic andesite in large homogeneous blocks. We have seen that the seismological activity is low, as well as the explosion related seismic activity. It is remarkable that we were able to connect only about 20\% of the observed noise peaks in the specific frequency region 1-10Hz to seismological, seismic or mine work related sources. 

\begin{figure}
\centering
\includegraphics[width=0.8\textwidth]{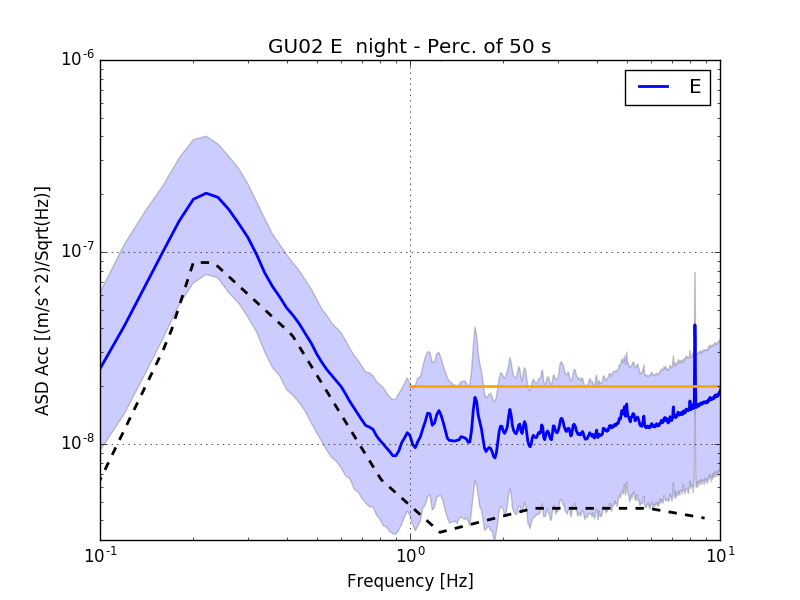}
\caption{\label{fig:2-week-PSD-HHE-night} Acceleration ASD values of the two-week period of GU02 seismometer (-400m), from the 50s long night percentiles in one of the horizontal directions. The dashed line is the NLNM curve of Peterson.}
\end{figure}

Considering the already established spectral and cumulative requirements by the ET Design Report \cite{ETdes11r} we can answer the previous question considering both the spectral and cumulative measures. From the point of view the spectral performance Figure \ref{fig:2-week-PSD-HHE-night} is instructive. Here the acceleration amplitude spectral density is shown for a period of two-week at a 404m deep location in the Gyöngyösoroszi mine at nightime (20:00-02:00CET each day). The spectral ET requirement, the so called Black Forest line is shown in brown and the upper and lower borders of the blue stripe and the middle blue line are the 90th, 10th and the 50th percentiles from the 50s averages of the data. Here the upper borderline, the 90th percentile indicates that the noise level is reliably under the ET requirement for 90\% of the chosen time periods. 

Regarding the cumulative measure  $rms_{2Hz} = 0.073 nm$ averaged for the horizontal directions and calculated from the 50s averages considering the modus with $0.1dB$ bins. 

\begin{figure}
\centering
\includegraphics[width=0.49\textwidth]{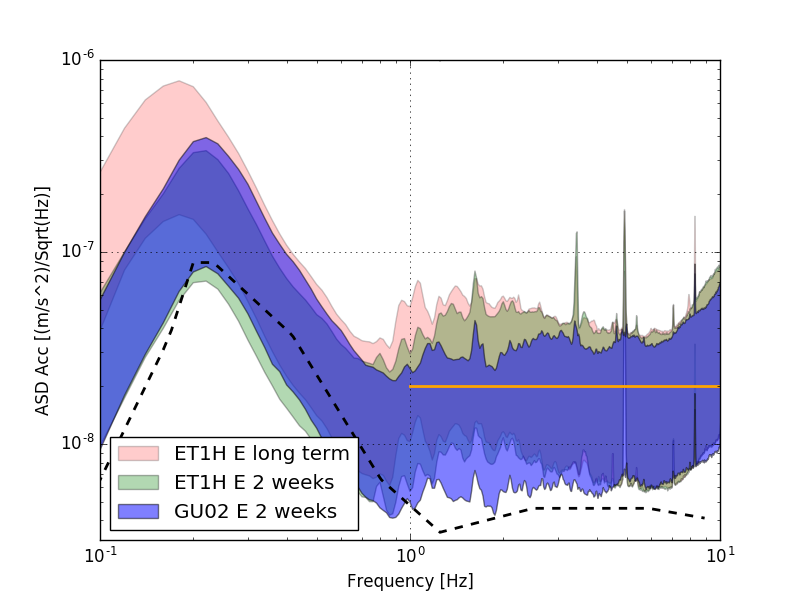}
\includegraphics[width=0.49\textwidth]{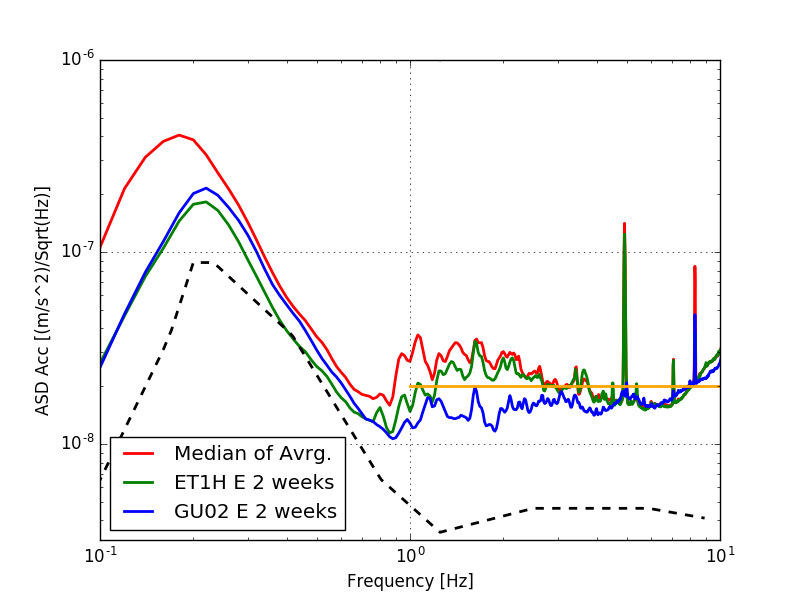}
\caption{\label{fig:osszvetes} Horizontal acceleration ASD values of the same two-week period of the GU02 seismometer (-400m) and the ETH1 seismometer (-88m) and the complete 600 days data of the  of the ETH1 station with, green and red colors, respectively. At the left hand side the stripes indicate the area between the 10 and 90 percentiles, at the right hand side the median  of the same data is shown with similar colors.}
\end{figure}

This is a kind of reliable optimal performance but not a typical one. How is it related to the average noise level and to the noise closer to the surface? Figure \ref{fig:osszvetes} can provide an impression.
At the left hand side the blue stripe indicates the the 10th-90th percentiles of the data from the same two-week of the same GU02 station considering the whole day measurements. The 90\% line is between $3-4\cdot10^{-8}\frac{m/s^{2}}{\sqrt{Hz}}$ during the whole period, which is not far from the Black Forest line value. Also remarkable that considering the data of our whole measurement campaign of almost two-year, indicated by the red stripe, the noise did not seem to increase. This is because in the Mátra there are no significant seasonal noise differences, as it was shown previously in detail. Nevertheless, there are differences, as it is more visible at the right hand side figure, on the medians of the previous data. What is remarkable here is that all three lines go close to the New Low Noise Level curve of Peterson down to 0.3Hz and  the convergence of the  three curves around 10Hz. It is not because of the lacking civilization noise, but because the median does not indicate the upper shifted noise level of the noisier half of the day. However, it indicates that the less noisier half remains calm for the whole year!
 
Regarding the cumulative characteristics, Table \ref{tab:rms_mod_comp} shows the {\it rms} calculated from the mode for the representative four cases seen on the acceleration ASD spectra, exactly as was in the short term studies \cite{ETdes11r,Bek13t,BekEta15a}. The corresponding median {\it rms} values may be higher or lower, depending on the noise strength distribution and on the exact calculation method.  It is remarkable that the 404m deep GU02 station measured the very same horizontal {\it rms} at the same place for a longer period. One can consider that value as representative in this depth for a long period, because of the seasonal independence of the noise in the MGGL, which is closer to the surface. It is also remarkable that the ongoing three shaft construction activity in 2017 was mostly in the vicinity of the ET1H station in MGGL.  

\begin{table}
\centering
\begin{tabular}{|c|c|c|c|c|}
\hline 
 & GU02 2w night & GU02 2w & ET1H 2w & ET1H whole\tabularnewline
\hline 
\hline 
$rms_{2Hz}$ [nm] & 0.073 & \boxed{0.083}  & 0.133 & 0.148 \tabularnewline
\hline 
\end{tabular}
\caption{\label{tab:rms_mod_comp} Mode related $rms_{2Hz}$ values (average horizontal) of the two-week measurements of the GU02 station (-404m) at night, and for whole period  and for the ET1H station  of MGGL (-88m) for the same two-week and for the whole {\dnum
} days. The night {\it rms} is calculated from 50s averages and the others from hourly averages. }
\end{table}

\section{Long term silence measures summary \label{sec:Conclusion2}}

In the previous sections we have surveyed and characterised the specific aspects of long term evaluation in order to find the best characterisation of the required ET silence. Our general observation is, that any performance measure can be suitable for site comparison as long as the calculation method is exactly the same and the noise distributions in time, strength and frequency are sufficiently similar. However, these performance measures are not the same from the point of view of ET requirements. From this point of view the percentiles seem to be the best spectral and cumulative characterisation, because they directly refer  operational conditions for the low frequency part of the ET. In particular the 90th percentile spectral distribution and the related {\it rms }seems to be a good candidate for filtering detection preventing, fatal noise levels. 

According to our previous analysis the characterization of long term noise data requires special consideration of the following aspects
\begin{enumerate}
\item The low noise estimation is distorted by short but intensive periods. These cannot be avoided and must be filtered. Lacking any objective criteria we suggest an intrinsic filtering performed by using the strength percentiles of the data. The 90th percentile indicates the noise level, below which ET can operate in 90\% of the time.
\item The typical noise level is best characterized by the 50th percentile, the median, for the ET purposes. The modus is unstable if the distribution of the data develops new peaks, which may happen in our case. 
\item The length of the intermediate averaging periods may be chosen according to ET operation requirements. In our study we have used mostly daily averaging, which is best for long term differences. 
\end{enumerate} 

We have seen that several seemingly minor aspects of the noise measures (e.g. the bin width of the modus calculations) may introduce different numbers and spectra emphasizing different properties of the overall noisiness. Long periods are more sensitive to these aspects than short ones.  

Also there are several further major aspects that can be related for the particular realisation of future underground gravitational wave detectors. For example if the design enables observation of signals below 2Hz, then the $rms_{2Hz}$ is an inadequate performance measure. 

Considering all these possible aspects it is very important to publish the raw seismological data for open evaluation and comparison. For the ET1H station of MGGL and for the GU02 station these are available.

\section{Acknowledgement}

The work was supported by the grants National Research, Development and Innovation Office \textendash{} NKFIH 116197(116375) NKFIH 124366(124366) and NKFIH 123815. The authors thank Péter Lévai and Géza Huba for the constant support and help, for Zoltán Zimborás and István János Kovács for valuable discussions. The first wonderful Mátra map is composed by Tamás Biró. Also the help and support of Árpád Váradi and Vilmos Rofrits from  Nitrokémia Zrt. and Róbert Weisz Róbert from GEO-FABER Zrt. is greatly acknowledged.

\bibliographystyle{unsrt}

\end{document}